\documentclass[aps,pra,twocolumn,,superscriptaddress,floatfix]{revtex4-1}
\usepackage[svgnames]{xcolor}
\definecolor{darkred}{rgb}{0.6, 0, 0}
\definecolor{darkgreen}{rgb}{0, 0.5, 0}
\usepackage[colorlinks=true,urlcolor = darkred,linkcolor=darkgreen,citecolor=darkred,bookmarks=false]{hyperref}
\usepackage{amsfonts,amsmath,amssymb}
\usepackage{graphicx}
\usepackage{dsfont}
\usepackage[normalem]{ulem}
\usepackage{enumerate}

\usepackage{tikz}
\usetikzlibrary{arrows,patterns,positioning,shadows}

\usepackage{pgfplots}

\def\ii{{\rm i}}
\newcommand{\dd}{{\rm d}}

\def\expect#1{\langle#1\rangle}

\begin{document}

\title{Universality classes of spin transport in one-dimensional isotropic magnets:\\
the onset of logarithmic anomalies}

\author{Jacopo De Nardis}
\affiliation{Department of Physics and Astronomy, University of Ghent, 
Krijgslaan 281, 9000 Gent, Belgium}

\author{Marko Medenjak}
\affiliation
{Institut de Physique Th\'eorique Philippe Meyer, \'Ecole Normale Sup\'erieure, \\ PSL University, Sorbonne Universit\'es, CNRS, 75005 Paris, France}

\author{Christoph Karrasch}
\affiliation{Technische Universit\"at Braunschweig, Institut f\"ur Mathematische Physik, Mendelssohnstra\ss e 3, 38106 Braunschweig, Germany}

\author{Enej Ilievski}
\affiliation{Faculty for Mathematics and Physics,
University of Ljubljana, Jadranska ulica 19, 1000 Ljubljana, Slovenia}
\affiliation{Institute for Theoretical Physics Amsterdam and Delta Institute for Theoretical Physics,
University of Amsterdam, Science Park 904, 1098 XH Amsterdam, The Netherlands}

\date{\today}

\begin{abstract}
We report a systematic study of finite-temperature spin transport in quantum and classical one-dimensional magnets with
isotropic spin interactions, including both integrable and non-integrable models.
Employing a phenomenological framework based on a generalized Burgers' equation in a time-dependent stochastic environment,
we identify four different universality classes of spin fluctuations. These comprise, aside from normal
spin diffusion, three types of superdiffusive transport: the KPZ universality class and two distinct types of anomalous diffusion
with multiplicative logarithmic corrections. Our predictions are supported by extensive numerical simulations on
various examples of quantum and classical chains. Contrary to common belief, we demonstrate
that even non-integrable spin chains can display a diverging spin diffusion constant at finite temperatures.
\end{abstract}


\maketitle
\paragraph*{\bf Introduction.} 

Obtaining a theoretical framework that is able to explain how macroscopic laws of transport emerge from the microscopic deterministic dynamics presents one of the central challenges of condensed matter physics.
This transcends purely academic interest, as many problems of quantum transport remain unresolved in the presence of strong interactions \cite{PhysRevLett.74.3253,RevModPhys.75.1085}. One viable strategy to improve our understanding of transport phenomena is to identify universality classes and study certain representative instances which can either be solved exactly, or at least
simulated numerically in an efficient manner \cite{Damle1998,PhysRevE.85.045201,PhysRevLett.70.1155,PhysRevLett.115.120602}.
In this respect interacting many-particle systems confined to one spatial dimension,
in the realms of both quantum and classical models, take a special role as they often exhibit anomalous features \cite{PhysRevLett.78.1896,Lepri2003}.
One of the prominent examples of a nonequilibrium universality class is given by the Kardar--Parisi--Zhang (KPZ)
equation \cite{KPZ} which is widespread in the area of growing one-dimensional interfaces \cite{Corwin2012,Takeuchi2018}.
The KPZ and L\'{e}vy universality classes which also occur in systems of classical particles can be understood in the scope of
the nonlinear fluctuating hydrodynamics \cite{Spohn1991,Prahofer2004,Spohn2014,Kulkarni2015,Popkov_2015}.

In recent years, the advent of the generalized hydrodynamics \cite{PhysRevX.6.041065,PhysRevLett.117.207201}, studies
of quantum chaos and its relation to transport \cite{Das2018,Gopalakrishnan2018,PhysRevX.9.041017,PhysRevLett.122.250603,1702.08894}, and of noisy quantum systems \cite{2001.04278,1912.08458,PhysRevLett.123.080601}, 
reinvigorated the field of transport laws in spin chain models.
In integrable quantum chains a closed-form universal expression for the conductivity matrix was found,
comprising both the Drude weights \cite{IN_Drude,Bulchandani2018,SciPostPhys.3.6.039,IN_Hubbard}
and diffusion constants \cite{DeNardis2018,Gopalakrishnan2018,DeNardis_SciPost,medenjak2019diffusion,doyon2019diffusion},
together with a myriad of other applications \cite{Bulchandani2017,PhysRevLett.120.045301,PhysRevLett.119.195301,1810.07170,Misguich2017,Misguich2019,Gamayun2019}.
This provided a coherent picture for earlier numerical results (see, e.g., \cite{Sirker2011,Karrasch2013,Karrasch2014,Steinigeweg2015,Steinigeweg2017} and references therein).
In spite of these developments, the discovery of superdiffusive spin transport and KPZ scaling (cf. Eq.~\eqref{eqn:KPZ_scaling})
in integrable spin chains with isotropic interactions, originally discovered numerically in the Heisenberg spin-$1/2$ chain
in \cite{PhysRevLett.106.220601,Ljubotina2017} and further
surveyed in \cite{Ilievski2018,GV2019,NMKI2019,DupontMoore2019,Vir2019}, came as a surprise.
Although recent numerical works \cite{Ljubotina2019,PhysRevE.100.042116,DupontMoore2019,Krajnik2019}, in combination with 
scaling arguments explaining the dynamical exponent \cite{GV2019}, constitute convincing evidence in support
of the KPZ universality, a rigorous analytical account of this phenomenon is still lacking.
Most recent studies suggest that anomalous spin transport occurs only in integrable systems invariant under non-Abelian
($SU(2)$ or $SO(3)$) Lie groups.
In contrast, normal spin diffusion is expected to be immediately recovered upon breaking
integrability \cite{DupontMoore2019,Krajnik2019}. The occurrence of normal diffusion in non-integrable symmetric chains was also 
suggested in the numerical study \cite{Bagchi2013} (see also \cite{Krajnik2019,Vir2019}), after a long-lasting controversy \cite{Muller1988,Gerling1990,Liu1991,Bonfim1992,Bohm1993,Srivastava1994}.
This perspective has been further reinforced in \cite{Vir2019} which proposes a phenomenological explanation of the KPZ-type 
superdiffusion. At this stage, two key questions remain unanswered:
\begin{enumerate}[(i)]
	\item  Do all rotationally invariant \textit{integrable} spin chains display superdiffusive spin transport?
	\item  Do all \textit{non-integrable} homogeneous spin chains display normal diffusive spin transport?
\end{enumerate}
To address these questions, we carry out a systematic study of magnetization transport in classical and quantum spin systems
in the \textit{non-magnetized sector} of thermal equilibrium states where the global rotational symmetry remains unbroken.
Aiming at complete classification of admissible transport laws, we build on a recent work by V. B. Bulchandani \cite{Vir2019} and 
devise a simple model that help us to single out two novel nonequilibrium universality classes of spin transport. We corroborate our findings with extensive numerical simulations of classical and quantum chains. \textit{Quite unexpectedly, we find that the
answers to questions (i) and (ii) are both negative}: the presence of global non-Abelian symmetries alongside integrability is \emph{not} a sufficient condition for an anomalous spin transport of the KPZ-type. Perhaps even
more surprisingly, despite the lack of integrability in the classical isotropic Heisenberg chain, the spin diffusion constant
is found to diverge logarithmically in time, thus refuting a widely held belief that non-integrable models cannot display infinite 
diffusion constants at finite temperatures.
From our perspective, the results summarized in diagram \ref{fig:diagram} should provide a comprehensive classification of magnetization
transport in homogeneous rotationally invariant quantum and classical spin models with short-range interactions.

We proceed by first introducing the formalism of an effective spin field theory which we relate to the KPZ equation in
a time-dependent noisy environment. To systematically test our predictions, we subsequently concentrate on a number of simple 
representative examples.

\paragraph*{\bf Spin-field theory of isotropic magnets.}

In order to capture various universality classes of spin superdiffusion, the task at hand is to devise an effective theory for finite-temperature magnetization dynamics in quantum and classical spin systems
with isotropic interactions valid on large spatio-temporal scales. Here we propose an effective description which is inherently classical in nature, by employing the continuum theory for 
a classical spin-field which is in turn treated within a hydrodynamic approximation. Microscopic details are included implicitly
through an appropriate phenomenological noise.

As our starting point we consider the most general form of a manifestly $SO(3)$-symmetric Hamiltonian equation of motion,
\begin{equation}
\vec{S}_{t} = -\vec{S}\times \frac{\delta H}{\delta \vec{S}}
= \mathcal{F}[\vec{S},\vec{S}_{x},\vec{S}_{xx},\ldots],\qquad |\vec{S}|=1,
\label{eqn:abstract_EOM}
\end{equation}
specified by some functional $\mathcal{F}$ involving scalar and vector products of the spin-field and derivatives thereof.
We can include classical lattice models as well, which are analyzed through their continuum counterparts.
To additionally incorporate quantum spin chains we first perform a mean-field average \footnote{For spin-$S$ quantum chains linear
in spin generators this is achieved by taking the expectation value on the $SU(2)$ spin-coherent states. Spin chains with nonlinear 
terms, where one encounters path-integral anomalies \cite{Wilson2011}, have to be excluded from our framework.} of the microscopic spin Hamiltonian.
It has to be stressed that such a correspondence cannot retain all \textit{quantitative} features of spin dynamics.
Nonetheless, we shall argue, in the spirit of \cite{Vir2019},  that correspondence is still meaningful to capture the correct
large-time transport behaviour.

The outlined effective theory applies in thermal equilibrium in the non-magnetized sector (i.e. at half-filling) where the global 
rotational invariance of the underlying invariant measure is unbroken. This is of paramount importance for the anomalous character of 
magnetization dynamics, as the addition of finite chemical potential (or external magnetic field), which dynamically breaks the
non-Abelian symmetry, leads to restoration of  normal spin diffusion (accompanied in integrable systems by a finite spin Drude weight \cite{PhysRevB.84.155125} or ballistic current). 
On large spatio-temporal scales, the evolution of a spin-field is accurately captured by a `hydrodynamic soft mode'
carrying a negligible energy density, which can be conveniently described in terms of two intrinsic geometric quantities,
curvature $\kappa = \big(\vec{S}_{x}\cdot \vec{S}_{x}\big)^{1/2}$ and torsion
$\tau = \kappa^{-2}\vec{S}\cdot (\vec{S}_{x}\times \vec{S}_{xx})$.
The soft mode pertains to long-wavelength ($k\sim \mathcal{O}(1/\ell)$) limit of the spin procession (about 
a distinguished axis fixed by the perturbation which breaks the gauge invariance, assumed subsequently to be the $z$-axis) at constant 
latitude $S^{z}=h$, with $S^{x}(x,t)\pm \ii S^{y}(x,t) = \sqrt{1-h^{2}}\exp{[\pm\ii(k\,x+w\,t)]}$, frequency $w(k)=-k^{2}h$ and 
dispersion $\mathcal{E}(k) = \tfrac{1}{2}k^{2}(1-h^2)$, which in terms of the curve describes helices with
constant (on scale $\sim \ell^{-1}$) $\kappa = \sqrt{1-h^{2}}k$ and $\tau = h\,k$ \footnote{While in the {Landau--Lifshitz} hierarchy 
of commuting flows helices are stationary states of the full nonlinear Hamiltonian dynamics, their fate in generic spin systems is 
less clear. We nonetheless expect they become stabilized by effective decoupling of the curvature field at large times.}.
Hydrodynamic modulation on a characteristic scale $\ell$, leading to scaling $\kappa \sim \tau \sim \mathcal{O}(1/\ell)$,
indicates that the energy density $\mathcal{E} = \kappa^2/2 \sim \ell^{-2}$ is suppressed compared to the dynamics of torsion $\tau$
(see also \cite{Vir2019}). Further noticing that energy fluctuations are diffusive (known to hold in both integrable and
non-integrable systems \cite{medenjak2019diffusion,Das2019}), they can be
effectively decoupled from the fluctuations of the torsion field provided the latter are superdiffusive.
Based on this we can conclude that torsion $\tau$ remains the only relevant scalar field at large times and that,
since $\tau \sim h$ at small $h$, the finite-temperature spin-spin fluctuations
$\langle \vec{S}(x,t) \cdot \vec{S}(0,0)\rangle$ are proportional to fluctuations of
the torsional mode $\langle \tau(x,t) \tau(0,0)\rangle$.
We shall assume that such a decoupling  mechanism holds generically for equations of the form \eqref{eqn:abstract_EOM}. 

\begin{figure}[t]
\centering

\begin{tikzpicture}[x=2 cm, y=-1 cm, node distance=0 cm,outer sep = 0pt]
\tikzstyle{nonlinearity}=[draw, thick, rectangle,  minimum height=0.5 cm, minimum width=2.4 cm,
fill=gray!10,anchor=south west, drop shadow, rounded corners]
\tikzstyle{noise}=[draw, thick, rectangle, minimum height=1.5 cm, minimum width=0.5 cm, fill=white,
anchor=north east, drop shadow, rounded corners]

\tikzstyle{class}=[rectangle,draw, thick, minimum width=2.4 cm, minimum height=1.5 cm,
anchor=north west,text centered,text width=5 em, drop shadow, rounded corners]
\tikzstyle{D}=[class,fill=green!10]
\tikzstyle{I2}=[class,fill=red!10]
\tikzstyle{I3}=[class,fill=yellow!10]
\tikzstyle{N2}=[class,fill=blue!10]
\node[nonlinearity,label={[yshift=0.1 cm]\small relevant}] (n2) at (1,7.9) {$n=2$};
\node[nonlinearity,label={[yshift=0.05 cm]\small marginal}] (n3) at (2.25,7.9) {$n=3$};
\node[nonlinearity,label={[yshift=0.1 cm]\small irrelevant}] (n4) at (3.5,7.9) {$n\geq 4$};
\node[noise] (integrable) at (0.9,8.1) {\rotatebox{90}{\scriptsize{$\zeta=0$}}};
\node (label1) at (0.5,8.9) {\rotatebox{90}{ballistic}};
\node[noise] (critical) at (0.9,9.7) {\rotatebox{90}{\scriptsize{$\zeta=1/2$}}};
\node (label2) at (0.5,10.5) {\rotatebox{90}{diffusive}};
\node[I2] at (1,8.1) {{\bf KPZ} $z=3/2$ \, $b=0$};
\node[N2] at (1,9.7) {${\rm {\bf D^{(2)}_{\rm log} }}$\\ $z=2$\, $b=2/3$};
\node[I3] at (2.25,8.1) {${\rm {\bf D^{(3)}_{\rm log } }}$\\ $z=2$\, $b=1/2$};
\node[D] at (2.25,9.7) {${\rm {\bf D }}$\\ $z=2$ \, $b=0$};
\node[D] at (3.5,8.1) {${\rm {\bf D}}$\\ $z=2$ \, $b=0$};
\node[D] at (3.5,9.7) {${\rm {\bf D}}$\\ $z=2$ \, $b=0$};
\end{tikzpicture}

\caption{Universality classes of spin dynamics in rotationally invariant spin systems.
The dynamical exponents $z$ and $b$ characterizing the time-asymptotic behavior of the spin dynamical correlations $\expect{\vec{S}(x,t) \cdot \vec{S}(0,0)}\sim t^{-1/z}\log^{-b}(t)$, as predicted by
the generalized noisy Burgers' equation for the hydrodynamic evolution of the torsion component 
$\tau_{t}+(\tau^{n}+D\tau_{x}+\sqrt{\gamma(t)}\,\eta)_{x}=0$ with a time-dependent 
noisy environment $\gamma(t)\sim t^{-\zeta}$.}
\label{fig:diagram}
\end{figure}

\paragraph*{\bf Generalized noisy Burgers' equation.} 

To account for thermal fluctuations we invoke the standard arguments of the nonlinear fluctuating hydrodynamics (NLFHD) 
\cite{Spohn2014}, where the microscopic degrees of freedom of the underlying Hamiltonian dynamics are effectively taken into account 
through an appropriate stochastic term and effective diffusion (i.e. dissipation). This brings us to
the \emph{generalized noisy Burgers' equation} of the form
\begin{equation}
\tau_{t} + (\tau^{n} + D\,\tau_{x} + \sqrt{\gamma}\,\eta)_{x} = 0,
\label{eqn:noisy_generalized_Burgers}
\end{equation}
where $D = \gamma/ \chi_\tau$ is the phenomenological diffusion constant, $\chi_\tau$ is static susceptibility of $\tau$, $\eta(x,t)$ a white noise with unit variance, $\gamma$ is the effective variance of the noisy environment and
parameter $n\geq 2$ specifies the degree of nonlinearity. From the general scaling relations
$\tau(x,t)\simeq t^{-1/2z}\,{\rm f}(x\,t^{-1/z})$ and $\expect{\tau(x,t)\tau(0,0)}\simeq t^{-1/z}\,{\rm g}(x\,t^{-1/z})$
(here and below the bracket refers to the average with respect to the canonical invariant measure) one however deduces that
$z=(n+1)/2$, which implies that nonlinearities of degree $n\geq 4$ (with $z>2$) are subdiffusive and thus irrelevant at large times.
Although NLFHD has no predictive power in this case, one generically expects to find normal spin transport.

The final key ingredient is to impose the structure of the noise, reflecting the nature of fluctuating modes which are relevant
on a hydrodynamic scale. In the spirit of conventional NLFHD (see e.g. \cite{Spohn2014}
for application to anharmonic chains), in the presence of long-lived ballistically propagating normal modes of 
Euler hydrodynamics, we adopt a time-independent white noise $\gamma =  \gamma_0$.
The same applies to integrable systems which exhibit extensively many local conserved fields, as previously suggested
in \cite{Vir2019}. On the contrary, generic spin systems do not support ballistic modes and excitations dissipate through 
the system. In this case the variance $\gamma$ is expected to obey the \emph{diffusive scaling} and decay with time
\begin{equation}\label{eq:scalingt}
\gamma \equiv \gamma(t) \sim \left(t_0/(t_0+t)\right)^{\zeta} \qquad {\rm with}\quad \zeta=1/2,
\end{equation}
where $t_0$ denotes an unknown model- and temperature-dependent scale.
The picture behind this is that fluctuations excited by the spatio-temporal variation of the chemical potential
should dissipate away diffusively as their density decays to zero with exponent $\zeta=1/2$.

For $n=2$, Eq.~\eqref{eqn:noisy_generalized_Burgers} is just the ordinary noisy Burgers' equation equivalent  to the KPZ equation, up to a change of variable \cite{KPZ}. A recent work \cite{1909.11557} examined the properties of such
KPZ equations with time-dependent noise term of the form \eqref{eq:scalingt}, finding non-universal large-time behavior for
$\zeta>1/2$ and universal KPZ dynamics (with modified dynamical exponents) for $\zeta \leq  1/2$.
Exactly at the `critical' point $\zeta = 1/2$, corresponding precisely to diffusive spreading of microscopic excitations,
in \cite{1909.11557} the authors deduce a modified diffusive scaling $x \sim t^{1/2}\log^{2/3}{(t/t_{0})}$ (for the particular case the scaling 
should be understood as a lower bound and not a rigorous statement \cite{PB}, as numerics are not able to distinguish slightly different exponents, see also additional numerical data in \cite{SM}).
Keeping this in mind, the statistics of spin fluctuations in this case is expected to exhibit a crossover from
an effective KPZ dynamics at short-intermediate times $t \simeq t_0$,
\begin{equation}
\expect{\vec{S} (x,t) \cdot \vec{S}(0,0)}\simeq t^{-2/3}f_{\rm KPZ}\big(\Gamma_{\rm KPZ}\,x\,t^{-2/3}\big),
\label{eqn:KPZ_scaling}
\end{equation}
to the asymptotic scaling of the form
\begin{equation}
\expect{\vec{S} (x,t) \cdot \vec{S}(0,0)}\simeq\frac{G\big(\Gamma_{G}\,x\,t^{-1/2}\log^{-2/3}{(t/t_{0})}\big)}{t^{1/2}\log^{2/3}{(t/t_{0})}},
\label{eqn:log_diffusive_scaling}
\end{equation}
with Gaussian profile $G(x)\simeq e^{-x^{2}}$ in the limit $t \gg t_0$.
The anomalous form \eqref{eqn:log_diffusive_scaling} implies a divergent behavior
$\mathfrak{D}(t)\sim [\log{(t)}]^{4/3}$.

\paragraph*{\bf Integrable isotropic magnets.}

Our main example is the Heisenberg continuous magnet $H^{(2)}=\tfrac{1}{2}\int \dd x\,\vec{S}_{x}\cdot \vec{S}_{x}$
(using standard notation $\vec{S}_{x} = \partial_x \vec{S}$ etc. for partial derivatives),
also known as the isotropic Landau--Lifshitz model \cite{Takhtajan1977,Faddeev1987},
which is a paradigmatic example of an \emph{integrable} classical field theory.
The time-evolution is governed by the nonlinear PDE $\vec{S}_{t} = F^{(2)}_{\rm LL} = \vec{S}\times \vec{S}_{xx}$.
This equation possesses an infinite family of Hamiltonians $H^{(n)}$ in involution,
e.g. $H^{(3)} =  \tfrac{1}{2}\int \dd x\,\vec{S}\cdot (\vec{S}_{x}\times \vec{S}_{xx})$ and
$H^{(4)} = -\tfrac{1}{2}\int \dd x \left[\vec{S}_{xx}\cdot \vec{S}_{xx} -
\tfrac{5}{4}\big( \vec{S}_{x}\cdot \vec{S}_{x} \big)^{2}\right]$.
Invoking the `decoupling hypothesis' and the phenomenological noisy environment, the torsional mode in each $H^{(n)}$ is
governed by the generalized Burgers' equations \eqref{eqn:noisy_generalized_Burgers}
with non-linearity of degree $n$ (see also \cite{SM} for the details).
Adopting a constant value for $\gamma$, the dynamics falls into the KPZ class at the lowest order $n=2$.
This is however no longer the case for $n>2$ where the quadratic nonlinearity is absent.
For $n\geq 4$ the non-linearity is dominated by diffusive processes and Hamiltonians $H^{(n>3)}$ thus
do not display any enhancement of normal diffusion.
The cubic $n=3$ case is however \emph{marginally irrelevant} in the dynamic RG sense. This type of nonlinearity
has been previously examined in the study of Toom interface \cite{Derrida1991,Devillard1992,Paczuski1992}
and argued to result in a logarithmic-type correction. We shall corroborate on this scenario later on.

\begin{figure}[t]
\centering
\includegraphics[width=0.9\columnwidth]{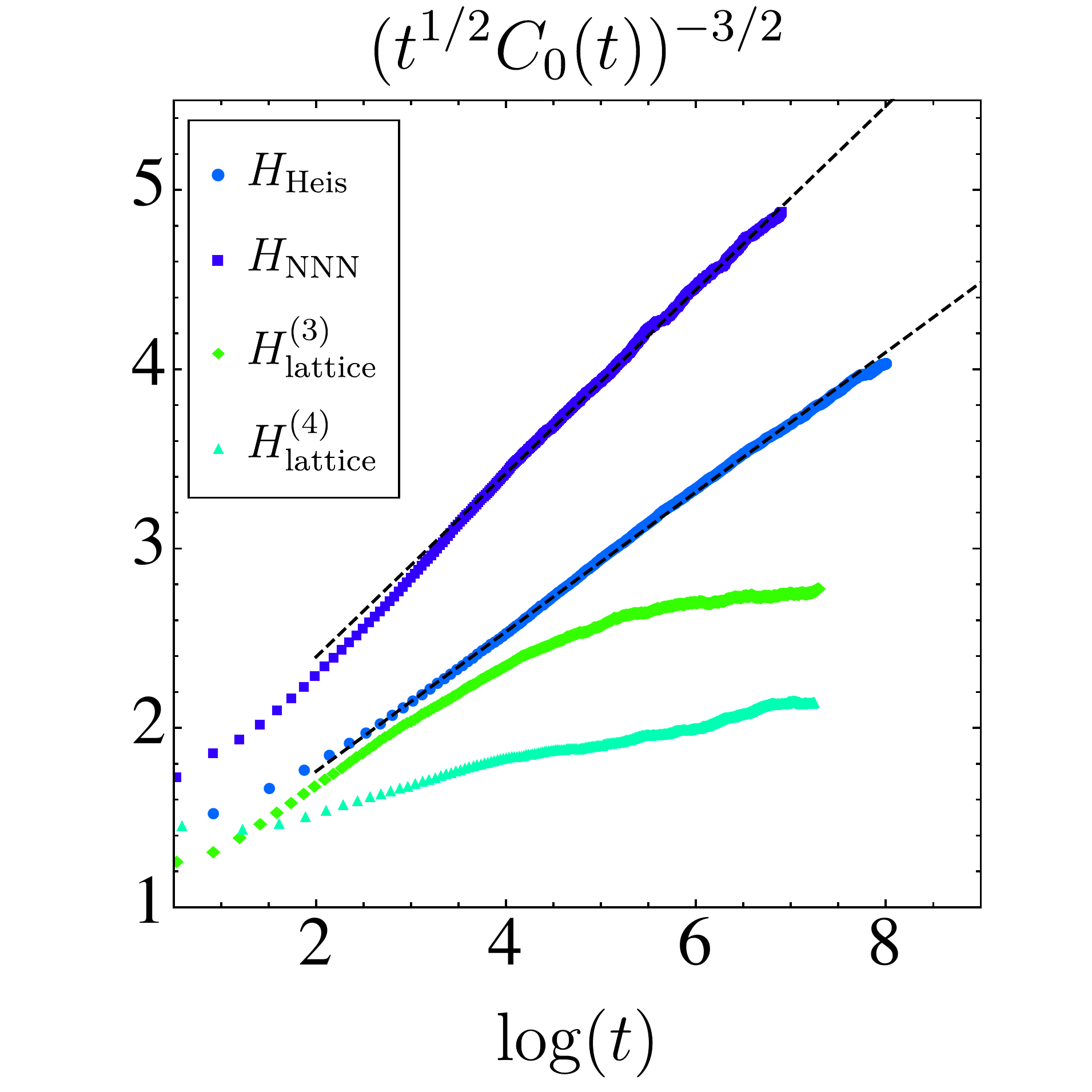}
\caption{Spin correlation function $C_0(t)= \langle S^z_{L/2}(t) S^z_{L/2}(0) \rangle$ plotted
as function $[t^{1/2}C_{0}(t)]^{-3/2}$ depending on $\log{(t)}$, shown for a few representative classical non-integrable lattice
spin models at infinite temperature.
The dashed curves are fitting lines with $\log{(t)}$. The Heisenberg and next-to-nearest neighbour Hamiltonians belong
to the ${\rm {\bf D }_{\rm log }^{(2)}}$ class, whereas the non-integrable lattice discretizations of $H^{(n)}$
(denoted by $H^{(n)}_{\rm lattice}$) exhibit normal diffusion ${\rm {\bf D}}$ for $n\geq 3$.}
\label{fig:log}
\end{figure}

\paragraph*{The quantum Heisenberg hierarchy.}
We proceed by examining the spin diffusion constant in integrable quantum spin-${\rm S}$ Heisenberg chains,
$\hat{H}^{(2)}=\sum_{j}\hat{\vec{S}}_{j}\cdot \hat{\vec{S}}_{j+1}$,
together with its hierarchy of local conservation laws $\hat{H}^{(n)}=\sum_{j}\hat{h}^{(n)}_{j}$ with $n$-site
densities $\hat{h}^{(n)}_{j}$ (hats denotes quantum operators). These models, we believe, exhaust all homogeneous
$SU(2)$-symmetric integrable quantum spin chains with short-range interactions (excluding cases symmetric
under higher-rank Lie groups).

Taking full advantage of quantum integrability and the underlying quasi-particle picture \cite{PhysRevX.6.041065,PhysRevLett.117.207201}, we have computed the exact spin diffusion constant, being
the dominant contribution to spin transport in the zero magnetization sector where the Drude weigh vanishes.
The details of this computations are spelled out in \cite{SM}, where we show that $\mathfrak{D}^{(n\geq 4)}<\infty$.
For the marginal case $n=3$ we however find a \emph{logarithmic divergence} $\mathfrak{D}^{(3)}(t)\sim \log{(t)}$.
This is our first example of a logarithmically enhanced diffusion, labelled by ${\rm {\bf D }_{\rm log }^{(3)}}$ in Fig.~\ref{fig:diagram}. As a proof of principle, the same conclusion can be independently reached by following the lines
of our phenomenological programme, using that the appropriate classical mean-field continuous limits of the Hamiltonians 
$\hat{H}^{(n)}$ are the higher Landau--Lifshitz Hamiltonians $H^{(n)}$ (cf. \cite{KMMZ2004,Bargheer2008}), which are
subsequently reduced to effective Burgers' equations of the form \eqref{eqn:noisy_generalized_Burgers}.
\begin{figure}[t]
\centering
\includegraphics[width=0.9\columnwidth]{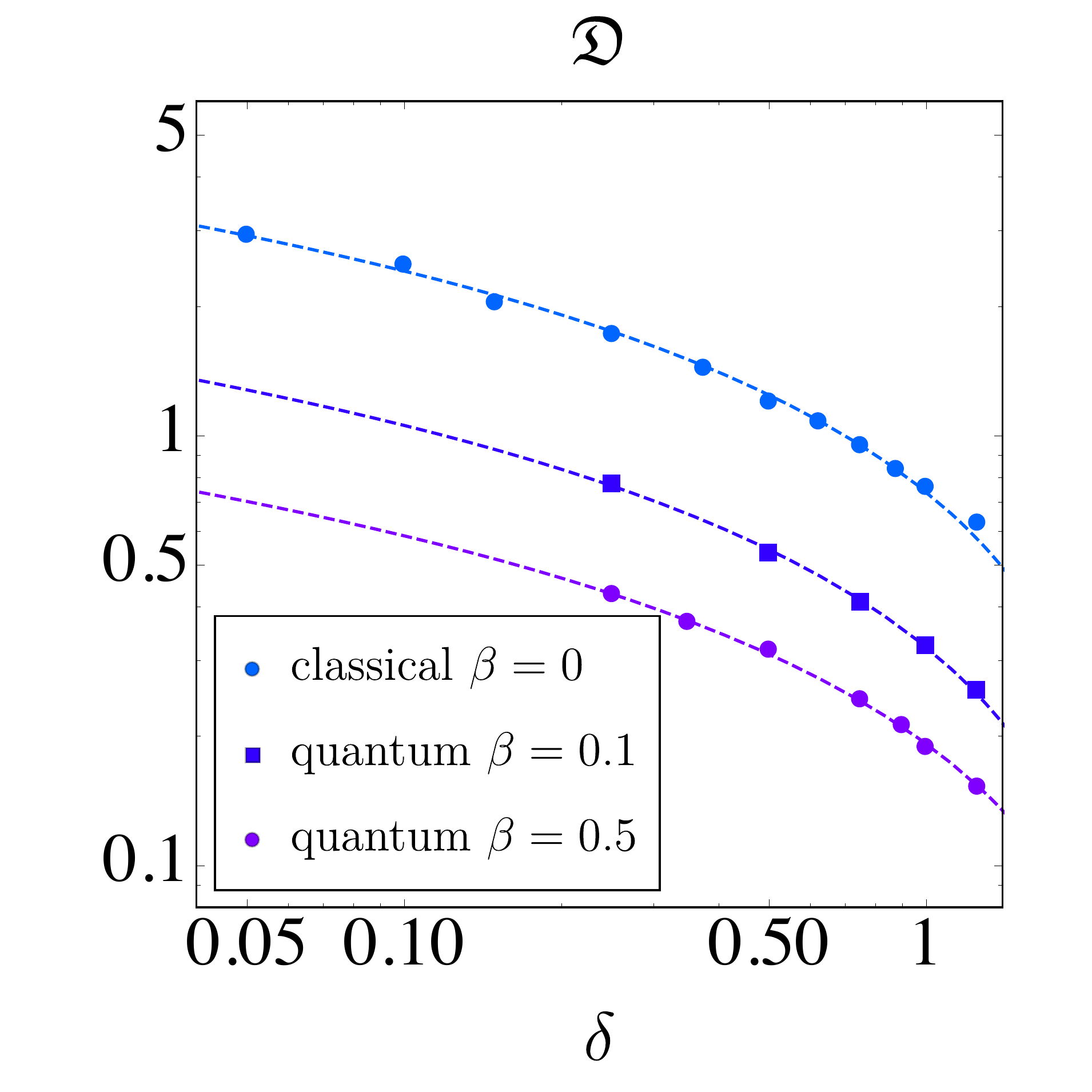}
\caption{Plot of the spin diffusion constant $\mathfrak{D}$ as function of the interaction anisotropy $\delta$
for the classical anisotropic chain $H_\delta$ and the quantum non-integrable spin-$1$ chain  $\hat{H}_\delta$  for two values of inverse temperature $\beta=1/T$. Dashed lines are fitting curves $\mathfrak{D}=-a^2\,\log{(c^2\,\delta)}$, with fitting parameters $a$ and $c$.}
\label{fig:delta}
\end{figure}

\paragraph*{\bf Non-integrable classical and quantum spin chains.} 
To demonstrate how non-integrable chains also fall in the classification scheme of Fig. \ref{fig:diagram},
we next examine the non-integrable classical Heisenberg chain $H_{\rm Heis}=\sum_{j}\vec{S}_{j}\cdot \vec{S}_{j+1}$,
where the most recent works align in favour of normal spin diffusion \cite{Bagchi2013} (see also \cite{Krajnik2019,Li2019,Vir2019}).
Performing numerical simulations with the method of \cite{Bagchi2013} (which conserves \textit{exactly} energy density and $|\vec{S}_j|$ at all times, see \cite{SM} for more details on the numerical simulations) and checking in addition the next-nearest neighbour scalar interaction, $H_{\rm NNN}=\sum_{j}\vec{S}_{j}\cdot \vec{S}_{j+1} + 0.8 \vec{S}_{j}\cdot \vec{S}_{j+2}$ along
with the \textit{non-integrable} lattice analogues of $H^{(3)}$ and $H^{(4)}$ flows (denoted by $H^{(n)}_{\rm lattice}$),
we computed the time-dependent on-site correlator
$C_{0}(t)\equiv \expect{S^{z}_{L/2}(t)S^{z}_{L/2}(0)}$ by flipping the centre spin $\vec{S}_{L/2}$ in a chain of length $L=10^3$ and average over an ensemble of $\mathcal{O}(10^{6})$ random configurations and over time, with different time-intervals
$\Delta t \leq 0.05$. The time-dependent diffusion constant $\mathfrak{D}(t)$ has been
extracted with aid of the diffusive ansatz $C_{0}(t)\!\sim\! [t \, \mathfrak{D}(t)]^{-1/2}$.
The numerical results, shown in Fig. \ref{fig:log}, confirm that both $H_{\rm Heis}$ and $H_{\rm NNN}$ are compatible
with effective Burgers' equation \eqref{eqn:noisy_generalized_Burgers} with $n=2$ and diffusive noise $\zeta=1/2$, yielding
(contrary to claims in refs.~\cite{Bagchi2013,Das2018}) a logarithmically divergent spin diffusion constant
(${\rm {\bf D }_{\rm log }^{(2)}}$ universality class in Fig.~\ref{fig:diagram}).
The continuous counterpart of non-integrable Hamiltonian $H^{(3)}_{\rm lattice}$ is instead described by a time-inhomogeneous
cubic $n=3$ Burgers' equation. Despite the lack of theoretical prediction for the late-time dynamics in this case,
our numerics suggests that spin dynamics is most likely purely diffusive.
Finally, the simulation of $H^{(4)}_{\rm lattice}$ nicely conforms with normal diffusion as expected from
Eq.~\eqref{eqn:noisy_generalized_Burgers} with a higher-degree of nonlinearity.

As an independent test, we additionally considered the \emph{anisotropic} model
$H_{\delta}=\sum_{j}\vec{S}_{j}\cdot \vec{S}_{j+1}+\delta\,\hat{S}^{z}_{j}\cdot \hat{S}^{z}_{j+1}$,
with interaction anisotropy $\delta$ acting as a `regulator' which restores normal diffusion, see also additional numerical data in \cite{SM}. By computing the diffusion constant  $\mathfrak{D}(\delta)$ and monitoring its value by approaching $\delta \to 0^{+}$,
we find behavior (cf. Fig.~\ref{fig:delta}) compatible with a mild divergence
\begin{equation}\label{eq:divergence}
\mathfrak{D}(\delta) \sim  |\log{(\delta)}\,|^d.
\end{equation}
with an exponent $d \approx 1$. A reliable extraction of transport coefficient in quantum spin chains is a very
demanding task obstructed by a rapid growth of entanglement entropy which renders tDMRG simulations with fixed bond dimension
\textit{uncertain} at large times. Despite inherent issues,
a recent numerical study of isotropic spin-${\rm S}$ chains \cite{DupontMoore2019}
concluded in favour of normal spin diffusion ($z=2$) in nonintegrable spin chains (${ S}\geq 1$).
Here we prefer to facilitate a direct comparison with classical spin chains. To this end, we carried out tDMRG simulations  \cite{white1992,Schollwoeck2011,Karrasch2012} of the quantum anisotropic spin-$1$ chain $\hat{H}_{\delta}$ and, restricting ourselves 
to only moderately small $\delta$, numerically extracted the spin diffusion constant
from the time-dependent DC conductivity with a diffusive tail $\sigma(t)\simeq (\chi/T)\mathfrak{D} + c\,t^{-1/2}$
with spin susceptibility $\chi$ and fitting parameters $\mathfrak{D}$ and $c$, see additional numerical data in \cite{SM}.
The data shown in Fig.~\ref{fig:delta} indicates that the spin dynamics in the non-integrable spin-$1$ chain mirrors that of its 
classical counterparts and thus experiences the same divergence \eqref{eq:divergence} (${\rm {\bf D }_{\rm log }^{(2)}}$ class). Notice that in the quantum integrable $S=1/2$ chain the divergence in the $\delta \to 0^+$ limit is different as it
diverges polynomially as $\mathfrak{D} \sim \delta^{-1/2}$ \cite{DeNardis_SciPost,Gopalakrishnan2018}, signalling the onset
of the KPZ dynamical exponent at $\delta=0$.

\paragraph*{\bf Conclusions.} 

We have proposed a phenomenological description of finite-temperature spin transport in one-dimensional quantum and
classical systems with isotropic interactions. We have predicted four different classes, including in particular two distinct types
of logarithmically enhanced diffusion which have not been previously disclosed in the context of many-body deterministic systems.
We \textit{conjecture} that in homogeneous $SU(2)$ or $SO(3)$ spin systems with short-range interactions this list is exhaustive (cf. Fig.~\ref{fig:diagram}). While the outlined approach can adequately capture the qualitative features of spin dynamic
(dynamical exponents and logarithmic corrections) it \textit{does not} give an access to exact values of transport coefficients
or couplings of the effective hydrodynamical equations \eqref{eqn:noisy_generalized_Burgers}.
Important refinements in this direction are left to future works.

Our findings shine some light on the puzzling observations in our previous work \cite{NMKI2019} which discusses
spin dynamics in the context of Haldane antiferromagnets. The observed short-time behavior with approximate
exponent $z=3/2$ (numerically detected also in \cite{Richter2019}), later argued in \cite{DupontMoore2019} to be merely a 
pronounced transient regime which crossovers into normal diffusion, indeed plays nicely with the expected transient scenario assisted 
by an effective time-dependent noise. Nonetheless, we argue now that despite the broken integrability the spin diffusion constant
does not saturate at asymptotically large times.
This can be reconciled with the predictions of \cite{NMKI2019} based on the effective low-energy quantum field theory
provided that the transient is regulated by a temperature-dependent time-scale diverging in the $T\to 0$ limit, as a consequence of 
the diverging effective lifetime $t_0$ of the microscopic degrees of freedom in \eqref{eq:scalingt}, despite different mechanisms resembling the situation in gapless one-dimensional systems \cite{Huang2013,Vasseur_2016}.


Finally,  It is reasonable to expect that the divergence \eqref{eq:divergence} could be seen in a real 
experimental setting and we hope that this can be successfully addressed in the near future.

\paragraph*{\bf Acknowledgement.}
We thank G. Barraquand, B. Bertini, V. Bulchandani, P. Le Doussal, \v{Z}. Krajnik, J. Moore, T. Prosen, and H. Spohn for
extensive and illuminating discussions,
and the organizers of the workshop ``Thermalization, Many-Body-Localization and Generalized Hydrodynamics''
at the ICTS Bangalore for hospitality, where parts of this work were carried out. J.D.N. is supported by the Research
Foundation Flanders (FWO). E.I. is supported by the Slovenian Research Agency (ARRS) under the Programme P1-0402
and VENI grant number 680-47-454 by the Netherlands Organisation for Scientific Research (NWO).
C.K. acknowledges support by the Deutsche Forschungsgemeinschaft through the Emmy Noether
program (KA 3360/2-1).


\bibliography{LogDiffusion}

\begin{thebibliography}{88}%
\makeatletter
\providecommand \@ifxundefined [1]{%
 \@ifx{#1\undefined}
}%
\providecommand \@ifnum [1]{%
 \ifnum #1\expandafter \@firstoftwo
 \else \expandafter \@secondoftwo
 \fi
}%
\providecommand \@ifx [1]{%
 \ifx #1\expandafter \@firstoftwo
 \else \expandafter \@secondoftwo
 \fi
}%
\providecommand \natexlab [1]{#1}%
\providecommand \enquote  [1]{``#1''}%
\providecommand \bibnamefont  [1]{#1}%
\providecommand \bibfnamefont [1]{#1}%
\providecommand \citenamefont [1]{#1}%
\providecommand \href@noop [0]{\@secondoftwo}%
\providecommand \href [0]{\begingroup \@sanitize@url \@href}%
\providecommand \@href[1]{\@@startlink{#1}\@@href}%
\providecommand \@@href[1]{\endgroup#1\@@endlink}%
\providecommand \@sanitize@url [0]{\catcode `\\12\catcode `\$12\catcode
  `\&12\catcode `\#12\catcode `\^12\catcode `\_12\catcode `\%12\relax}%
\providecommand \@@startlink[1]{}%
\providecommand \@@endlink[0]{}%
\providecommand \url  [0]{\begingroup\@sanitize@url \@url }%
\providecommand \@url [1]{\endgroup\@href {#1}{\urlprefix }}%
\providecommand \urlprefix  [0]{URL }%
\providecommand \Eprint [0]{\href }%
\providecommand \doibase [0]{http://dx.doi.org/}%
\providecommand \selectlanguage [0]{\@gobble}%
\providecommand \bibinfo  [0]{\@secondoftwo}%
\providecommand \bibfield  [0]{\@secondoftwo}%
\providecommand \translation [1]{[#1]}%
\providecommand \BibitemOpen [0]{}%
\providecommand \bibitemStop [0]{}%
\providecommand \bibitemNoStop [0]{.\EOS\space}%
\providecommand \EOS [0]{\spacefactor3000\relax}%
\providecommand \BibitemShut  [1]{\csname bibitem#1\endcsname}%
\let\auto@bib@innerbib\@empty
\bibitem [{\citenamefont {Emery}\ and\ \citenamefont
  {Kivelson}(1995)}]{PhysRevLett.74.3253}%
  \BibitemOpen
  \bibfield  {author} {\bibinfo {author} {\bibfnamefont {V.~J.}\ \bibnamefont
  {Emery}}\ and\ \bibinfo {author} {\bibfnamefont {S.~A.}\ \bibnamefont
  {Kivelson}},\ }\href {\doibase 10.1103/PhysRevLett.74.3253} {\bibfield
  {journal} {\bibinfo  {journal} {Phys. Rev. Lett.}\ }\textbf {\bibinfo
  {volume} {74}},\ \bibinfo {pages} {3253} (\bibinfo {year}
  {1995})}\BibitemShut {NoStop}%
\bibitem [{\citenamefont {Gunnarsson}\ \emph {et~al.}(2003)\citenamefont
  {Gunnarsson}, \citenamefont {Calandra},\ and\ \citenamefont
  {Han}}]{RevModPhys.75.1085}%
  \BibitemOpen
  \bibfield  {author} {\bibinfo {author} {\bibfnamefont {O.}~\bibnamefont
  {Gunnarsson}}, \bibinfo {author} {\bibfnamefont {M.}~\bibnamefont
  {Calandra}}, \ and\ \bibinfo {author} {\bibfnamefont {J.~E.}\ \bibnamefont
  {Han}},\ }\href {\doibase 10.1103/RevModPhys.75.1085} {\bibfield  {journal}
  {\bibinfo  {journal} {Rev. Mod. Phys.}\ }\textbf {\bibinfo {volume} {75}},\
  \bibinfo {pages} {1085} (\bibinfo {year} {2003})}\BibitemShut {NoStop}%
\bibitem [{\citenamefont {Damle}\ and\ \citenamefont
  {Sachdev}(1998)}]{Damle1998}%
  \BibitemOpen
  \bibfield  {author} {\bibinfo {author} {\bibfnamefont {K.}~\bibnamefont
  {Damle}}\ and\ \bibinfo {author} {\bibfnamefont {S.}~\bibnamefont
  {Sachdev}},\ }\href {\doibase 10.1103/physrevb.57.8307} {\bibfield  {journal}
  {\bibinfo  {journal} {Physical Review B}\ }\textbf {\bibinfo {volume} {57}},\
  \bibinfo {pages} {8307} (\bibinfo {year} {1998})}\BibitemShut {NoStop}%
\bibitem [{\citenamefont {Berkolaiko}\ and\ \citenamefont
  {Kuipers}(2012)}]{PhysRevE.85.045201}%
  \BibitemOpen
  \bibfield  {author} {\bibinfo {author} {\bibfnamefont {G.}~\bibnamefont
  {Berkolaiko}}\ and\ \bibinfo {author} {\bibfnamefont {J.}~\bibnamefont
  {Kuipers}},\ }\href {\doibase 10.1103/PhysRevE.85.045201} {\bibfield
  {journal} {\bibinfo  {journal} {Phys. Rev. E}\ }\textbf {\bibinfo {volume}
  {85}},\ \bibinfo {pages} {045201} (\bibinfo {year} {2012})}\BibitemShut
  {NoStop}%
\bibitem [{\citenamefont {Beenakker}(1993)}]{PhysRevLett.70.1155}%
  \BibitemOpen
  \bibfield  {author} {\bibinfo {author} {\bibfnamefont {C.~W.~J.}\
  \bibnamefont {Beenakker}},\ }\href {\doibase 10.1103/PhysRevLett.70.1155}
  {\bibfield  {journal} {\bibinfo  {journal} {Phys. Rev. Lett.}\ }\textbf
  {\bibinfo {volume} {70}},\ \bibinfo {pages} {1155} (\bibinfo {year}
  {1993})}\BibitemShut {NoStop}%
\bibitem [{\citenamefont {Kulvelis}\ \emph {et~al.}(2015)\citenamefont
  {Kulvelis}, \citenamefont {Dolgushev},\ and\ \citenamefont
  {M\"ulken}}]{PhysRevLett.115.120602}%
  \BibitemOpen
  \bibfield  {author} {\bibinfo {author} {\bibfnamefont {N.}~\bibnamefont
  {Kulvelis}}, \bibinfo {author} {\bibfnamefont {M.}~\bibnamefont {Dolgushev}},
  \ and\ \bibinfo {author} {\bibfnamefont {O.}~\bibnamefont {M\"ulken}},\
  }\href {\doibase 10.1103/PhysRevLett.115.120602} {\bibfield  {journal}
  {\bibinfo  {journal} {Phys. Rev. Lett.}\ }\textbf {\bibinfo {volume} {115}},\
  \bibinfo {pages} {120602} (\bibinfo {year} {2015})}\BibitemShut {NoStop}%
\bibitem [{\citenamefont {Lepri}\ \emph {et~al.}(1997)\citenamefont {Lepri},
  \citenamefont {Livi},\ and\ \citenamefont {Politi}}]{PhysRevLett.78.1896}%
  \BibitemOpen
  \bibfield  {author} {\bibinfo {author} {\bibfnamefont {S.}~\bibnamefont
  {Lepri}}, \bibinfo {author} {\bibfnamefont {R.}~\bibnamefont {Livi}}, \ and\
  \bibinfo {author} {\bibfnamefont {A.}~\bibnamefont {Politi}},\ }\href
  {\doibase 10.1103/PhysRevLett.78.1896} {\bibfield  {journal} {\bibinfo
  {journal} {Phys. Rev. Lett.}\ }\textbf {\bibinfo {volume} {78}},\ \bibinfo
  {pages} {1896} (\bibinfo {year} {1997})}\BibitemShut {NoStop}%
\bibitem [{\citenamefont {Lepri}(2003)}]{Lepri2003}%
  \BibitemOpen
  \bibfield  {author} {\bibinfo {author} {\bibfnamefont {S.}~\bibnamefont
  {Lepri}},\ }\href {\doibase 10.1016/s0370-1573(02)00558-6} {\bibfield
  {journal} {\bibinfo  {journal} {Physics Reports}\ }\textbf {\bibinfo {volume}
  {377}},\ \bibinfo {pages} {1} (\bibinfo {year} {2003})}\BibitemShut {NoStop}%
\bibitem [{\citenamefont {Kardar}\ \emph {et~al.}(1986)\citenamefont {Kardar},
  \citenamefont {Parisi},\ and\ \citenamefont {Zhang}}]{KPZ}%
  \BibitemOpen
  \bibfield  {author} {\bibinfo {author} {\bibfnamefont {M.}~\bibnamefont
  {Kardar}}, \bibinfo {author} {\bibfnamefont {G.}~\bibnamefont {Parisi}}, \
  and\ \bibinfo {author} {\bibfnamefont {Y.-C.}\ \bibnamefont {Zhang}},\ }\href
  {\doibase 10.1103/PhysRevLett.56.889} {\bibfield  {journal} {\bibinfo
  {journal} {Phys. Rev. Lett.}\ }\textbf {\bibinfo {volume} {56}},\ \bibinfo
  {pages} {889} (\bibinfo {year} {1986})}\BibitemShut {NoStop}%
\bibitem [{\citenamefont {Corwin}(2012)}]{Corwin2012}%
  \BibitemOpen
  \bibfield  {author} {\bibinfo {author} {\bibfnamefont {I.}~\bibnamefont
  {Corwin}},\ }\href {\doibase 10.1142/s2010326311300014} {\bibfield  {journal}
  {\bibinfo  {journal} {Random Matrices: Theory and Applications}\ }\textbf
  {\bibinfo {volume} {01}},\ \bibinfo {pages} {1130001} (\bibinfo {year}
  {2012})}\BibitemShut {NoStop}%
\bibitem [{\citenamefont {Takeuchi}(2018)}]{Takeuchi2018}%
  \BibitemOpen
  \bibfield  {author} {\bibinfo {author} {\bibfnamefont {K.~A.}\ \bibnamefont
  {Takeuchi}},\ }\href {\doibase 10.1016/j.physa.2018.03.009} {\bibfield
  {journal} {\bibinfo  {journal} {Physica A: Statistical Mechanics and its
  Applications}\ }\textbf {\bibinfo {volume} {504}},\ \bibinfo {pages} {77}
  (\bibinfo {year} {2018})}\BibitemShut {NoStop}%
\bibitem [{\citenamefont {Spohn}(1991)}]{Spohn1991}%
  \BibitemOpen
  \bibfield  {author} {\bibinfo {author} {\bibfnamefont {H.}~\bibnamefont
  {Spohn}},\ }\href {\doibase 10.1007/978-3-642-84371-6} {\emph {\bibinfo
  {title} {Large Scale Dynamics of Interacting Particles}}}\ (\bibinfo
  {publisher} {Springer Berlin Heidelberg},\ \bibinfo {year}
  {1991})\BibitemShut {NoStop}%
\bibitem [{\citenamefont {Pr\"{a}hofer}\ and\ \citenamefont
  {Spohn}(2004)}]{Prahofer2004}%
  \BibitemOpen
  \bibfield  {author} {\bibinfo {author} {\bibfnamefont {M.}~\bibnamefont
  {Pr\"{a}hofer}}\ and\ \bibinfo {author} {\bibfnamefont {H.}~\bibnamefont
  {Spohn}},\ }\href {\doibase 10.1023/b:joss.0000019810.21828.fc} {\bibfield
  {journal} {\bibinfo  {journal} {Journal of Statistical Physics}\ }\textbf
  {\bibinfo {volume} {115}},\ \bibinfo {pages} {255} (\bibinfo {year}
  {2004})}\BibitemShut {NoStop}%
\bibitem [{\citenamefont {Spohn}(2014)}]{Spohn2014}%
  \BibitemOpen
  \bibfield  {author} {\bibinfo {author} {\bibfnamefont {H.}~\bibnamefont
  {Spohn}},\ }\href {\doibase 10.1007/s10955-014-0933-y} {\bibfield  {journal}
  {\bibinfo  {journal} {Journal of Statistical Physics}\ }\textbf {\bibinfo
  {volume} {154}},\ \bibinfo {pages} {1191} (\bibinfo {year}
  {2014})}\BibitemShut {NoStop}%
\bibitem [{\citenamefont {Kulkarni}\ \emph {et~al.}(2015)\citenamefont
  {Kulkarni}, \citenamefont {Huse},\ and\ \citenamefont
  {Spohn}}]{Kulkarni2015}%
  \BibitemOpen
  \bibfield  {author} {\bibinfo {author} {\bibfnamefont {M.}~\bibnamefont
  {Kulkarni}}, \bibinfo {author} {\bibfnamefont {D.~A.}\ \bibnamefont {Huse}},
  \ and\ \bibinfo {author} {\bibfnamefont {H.}~\bibnamefont {Spohn}},\ }\href
  {\doibase 10.1103/physreva.92.043612} {\bibfield  {journal} {\bibinfo
  {journal} {Physical Review A}\ }\textbf {\bibinfo {volume} {92}} (\bibinfo
  {year} {2015}),\ 10.1103/physreva.92.043612}\BibitemShut {NoStop}%
\bibitem [{\citenamefont {Popkov}\ \emph {et~al.}(2015)\citenamefont {Popkov},
  \citenamefont {Schadschneider}, \citenamefont {Schmidt},\ and\ \citenamefont
  {Sch\"{u}tz}}]{Popkov_2015}%
  \BibitemOpen
  \bibfield  {author} {\bibinfo {author} {\bibfnamefont {V.}~\bibnamefont
  {Popkov}}, \bibinfo {author} {\bibfnamefont {A.}~\bibnamefont
  {Schadschneider}}, \bibinfo {author} {\bibfnamefont {J.}~\bibnamefont
  {Schmidt}}, \ and\ \bibinfo {author} {\bibfnamefont {G.~M.}\ \bibnamefont
  {Sch\"{u}tz}},\ }\href {\doibase 10.1073/pnas.1512261112} {\bibfield
  {journal} {\bibinfo  {journal} {Proceedings of the National Academy of
  Sciences}\ }\textbf {\bibinfo {volume} {112}},\ \bibinfo {pages} {12645}
  (\bibinfo {year} {2015})}\BibitemShut {NoStop}%
\bibitem [{\citenamefont {Castro-Alvaredo}\ \emph {et~al.}(2016)\citenamefont
  {Castro-Alvaredo}, \citenamefont {Doyon},\ and\ \citenamefont
  {Yoshimura}}]{PhysRevX.6.041065}%
  \BibitemOpen
  \bibfield  {author} {\bibinfo {author} {\bibfnamefont {O.~A.}\ \bibnamefont
  {Castro-Alvaredo}}, \bibinfo {author} {\bibfnamefont {B.}~\bibnamefont
  {Doyon}}, \ and\ \bibinfo {author} {\bibfnamefont {T.}~\bibnamefont
  {Yoshimura}},\ }\href {\doibase 10.1103/PhysRevX.6.041065} {\bibfield
  {journal} {\bibinfo  {journal} {Phys. Rev. X}\ }\textbf {\bibinfo {volume}
  {6}},\ \bibinfo {pages} {041065} (\bibinfo {year} {2016})}\BibitemShut
  {NoStop}%
\bibitem [{\citenamefont {Bertini}\ \emph {et~al.}(2016)\citenamefont
  {Bertini}, \citenamefont {Collura}, \citenamefont {{De Nardis}},\ and\
  \citenamefont {Fagotti}}]{PhysRevLett.117.207201}%
  \BibitemOpen
  \bibfield  {author} {\bibinfo {author} {\bibfnamefont {B.}~\bibnamefont
  {Bertini}}, \bibinfo {author} {\bibfnamefont {M.}~\bibnamefont {Collura}},
  \bibinfo {author} {\bibfnamefont {J.}~\bibnamefont {{De Nardis}}}, \ and\
  \bibinfo {author} {\bibfnamefont {M.}~\bibnamefont {Fagotti}},\ }\href
  {\doibase 10.1103/PhysRevLett.117.207201} {\bibfield  {journal} {\bibinfo
  {journal} {Phys. Rev. Lett.}\ }\textbf {\bibinfo {volume} {117}},\ \bibinfo
  {pages} {207201} (\bibinfo {year} {2016})}\BibitemShut {NoStop}%
\bibitem [{\citenamefont {Das}\ \emph {et~al.}(2018{\natexlab{a}})\citenamefont
  {Das}, \citenamefont {Chakrabarty}, \citenamefont {Dhar}, \citenamefont
  {Kundu}, \citenamefont {Huse}, \citenamefont {Moessner}, \citenamefont
  {Ray},\ and\ \citenamefont {Bhattacharjee}}]{Das2018}%
  \BibitemOpen
  \bibfield  {author} {\bibinfo {author} {\bibfnamefont {A.}~\bibnamefont
  {Das}}, \bibinfo {author} {\bibfnamefont {S.}~\bibnamefont {Chakrabarty}},
  \bibinfo {author} {\bibfnamefont {A.}~\bibnamefont {Dhar}}, \bibinfo {author}
  {\bibfnamefont {A.}~\bibnamefont {Kundu}}, \bibinfo {author} {\bibfnamefont
  {D.~A.}\ \bibnamefont {Huse}}, \bibinfo {author} {\bibfnamefont
  {R.}~\bibnamefont {Moessner}}, \bibinfo {author} {\bibfnamefont {S.~S.}\
  \bibnamefont {Ray}}, \ and\ \bibinfo {author} {\bibfnamefont
  {S.}~\bibnamefont {Bhattacharjee}},\ }\href {\doibase
  10.1103/PhysRevLett.121.024101} {\bibfield  {journal} {\bibinfo  {journal}
  {Phys. Rev. Lett.}\ }\textbf {\bibinfo {volume} {121}},\ \bibinfo {pages}
  {024101} (\bibinfo {year} {2018}{\natexlab{a}})}\BibitemShut {NoStop}%
\bibitem [{\citenamefont {Gopalakrishnan}\ \emph {et~al.}(2018)\citenamefont
  {Gopalakrishnan}, \citenamefont {Huse}, \citenamefont {Khemani},\ and\
  \citenamefont {Vasseur}}]{Gopalakrishnan2018}%
  \BibitemOpen
  \bibfield  {author} {\bibinfo {author} {\bibfnamefont {S.}~\bibnamefont
  {Gopalakrishnan}}, \bibinfo {author} {\bibfnamefont {D.~A.}\ \bibnamefont
  {Huse}}, \bibinfo {author} {\bibfnamefont {V.}~\bibnamefont {Khemani}}, \
  and\ \bibinfo {author} {\bibfnamefont {R.}~\bibnamefont {Vasseur}},\ }\href
  {\doibase 10.1103/physrevb.98.220303} {\bibfield  {journal} {\bibinfo
  {journal} {Physical Review B}\ }\textbf {\bibinfo {volume} {98}} (\bibinfo
  {year} {2018}),\ 10.1103/physrevb.98.220303}\BibitemShut {NoStop}%
\bibitem [{\citenamefont {Parker}\ \emph {et~al.}(2019)\citenamefont {Parker},
  \citenamefont {Cao}, \citenamefont {Avdoshkin}, \citenamefont {Scaffidi},\
  and\ \citenamefont {Altman}}]{PhysRevX.9.041017}%
  \BibitemOpen
  \bibfield  {author} {\bibinfo {author} {\bibfnamefont {D.~E.}\ \bibnamefont
  {Parker}}, \bibinfo {author} {\bibfnamefont {X.}~\bibnamefont {Cao}},
  \bibinfo {author} {\bibfnamefont {A.}~\bibnamefont {Avdoshkin}}, \bibinfo
  {author} {\bibfnamefont {T.}~\bibnamefont {Scaffidi}}, \ and\ \bibinfo
  {author} {\bibfnamefont {E.}~\bibnamefont {Altman}},\ }\href {\doibase
  10.1103/PhysRevX.9.041017} {\bibfield  {journal} {\bibinfo  {journal} {Phys.
  Rev. X}\ }\textbf {\bibinfo {volume} {9}},\ \bibinfo {pages} {041017}
  (\bibinfo {year} {2019})}\BibitemShut {NoStop}%
\bibitem [{\citenamefont {Alba}\ \emph {et~al.}(2019)\citenamefont {Alba},
  \citenamefont {Dubail},\ and\ \citenamefont
  {Medenjak}}]{PhysRevLett.122.250603}%
  \BibitemOpen
  \bibfield  {author} {\bibinfo {author} {\bibfnamefont {V.}~\bibnamefont
  {Alba}}, \bibinfo {author} {\bibfnamefont {J.}~\bibnamefont {Dubail}}, \ and\
  \bibinfo {author} {\bibfnamefont {M.}~\bibnamefont {Medenjak}},\ }\href
  {\doibase 10.1103/PhysRevLett.122.250603} {\bibfield  {journal} {\bibinfo
  {journal} {Phys. Rev. Lett.}\ }\textbf {\bibinfo {volume} {122}},\ \bibinfo
  {pages} {250603} (\bibinfo {year} {2019})}\BibitemShut {NoStop}%
\bibitem [{\citenamefont {Leviatan}\ \emph {et~al.}(2017)\citenamefont
  {Leviatan}, \citenamefont {Pollmann}, \citenamefont {Bardarson},
  \citenamefont {Huse},\ and\ \citenamefont {Altman}}]{1702.08894}%
  \BibitemOpen
  \bibfield  {author} {\bibinfo {author} {\bibfnamefont {E.}~\bibnamefont
  {Leviatan}}, \bibinfo {author} {\bibfnamefont {F.}~\bibnamefont {Pollmann}},
  \bibinfo {author} {\bibfnamefont {J.~H.}\ \bibnamefont {Bardarson}}, \bibinfo
  {author} {\bibfnamefont {D.~A.}\ \bibnamefont {Huse}}, \ and\ \bibinfo
  {author} {\bibfnamefont {E.}~\bibnamefont {Altman}},\ }\href@noop {} {\
  (\bibinfo {year} {2017})},\ \Eprint {http://arxiv.org/abs/arXiv:1702.08894}
  {arXiv:1702.08894} \BibitemShut {NoStop}%
\bibitem [{\citenamefont {Jin}\ \emph {et~al.}(2020)\citenamefont {Jin},
  \citenamefont {Krajenbrink},\ and\ \citenamefont {Bernard}}]{2001.04278}%
  \BibitemOpen
  \bibfield  {author} {\bibinfo {author} {\bibfnamefont {T.}~\bibnamefont
  {Jin}}, \bibinfo {author} {\bibfnamefont {A.}~\bibnamefont {Krajenbrink}}, \
  and\ \bibinfo {author} {\bibfnamefont {D.}~\bibnamefont {Bernard}},\
  }\href@noop {} {\  (\bibinfo {year} {2020})},\ \Eprint
  {http://arxiv.org/abs/arXiv:2001.04278} {arXiv:2001.04278} \BibitemShut
  {NoStop}%
\bibitem [{\citenamefont {Bernard}\ and\ \citenamefont
  {Doussal}(2019)}]{1912.08458}%
  \BibitemOpen
  \bibfield  {author} {\bibinfo {author} {\bibfnamefont {D.}~\bibnamefont
  {Bernard}}\ and\ \bibinfo {author} {\bibfnamefont {P.~L.}\ \bibnamefont
  {Doussal}},\ }\href@noop {} {\  (\bibinfo {year} {2019})},\ \Eprint
  {http://arxiv.org/abs/arXiv:1912.08458} {arXiv:1912.08458} \BibitemShut
  {NoStop}%
\bibitem [{\citenamefont {Bernard}\ and\ \citenamefont
  {Jin}(2019)}]{PhysRevLett.123.080601}%
  \BibitemOpen
  \bibfield  {author} {\bibinfo {author} {\bibfnamefont {D.}~\bibnamefont
  {Bernard}}\ and\ \bibinfo {author} {\bibfnamefont {T.}~\bibnamefont {Jin}},\
  }\href {\doibase 10.1103/PhysRevLett.123.080601} {\bibfield  {journal}
  {\bibinfo  {journal} {Phys. Rev. Lett.}\ }\textbf {\bibinfo {volume} {123}},\
  \bibinfo {pages} {080601} (\bibinfo {year} {2019})}\BibitemShut {NoStop}%
\bibitem [{\citenamefont {Ilievski}\ and\ \citenamefont {{De
  Nardis}}(2017{\natexlab{a}})}]{IN_Drude}%
  \BibitemOpen
  \bibfield  {author} {\bibinfo {author} {\bibfnamefont {E.}~\bibnamefont
  {Ilievski}}\ and\ \bibinfo {author} {\bibfnamefont {J.}~\bibnamefont {{De
  Nardis}}},\ }\href {\doibase 10.1103/physrevlett.119.020602} {\bibfield
  {journal} {\bibinfo  {journal} {Physical Review Letters}\ }\textbf {\bibinfo
  {volume} {119}} (\bibinfo {year} {2017}{\natexlab{a}}),\
  10.1103/physrevlett.119.020602}\BibitemShut {NoStop}%
\bibitem [{\citenamefont {Bulchandani}\ \emph {et~al.}(2018)\citenamefont
  {Bulchandani}, \citenamefont {Vasseur}, \citenamefont {Karrasch},\ and\
  \citenamefont {Moore}}]{Bulchandani2018}%
  \BibitemOpen
  \bibfield  {author} {\bibinfo {author} {\bibfnamefont {V.~B.}\ \bibnamefont
  {Bulchandani}}, \bibinfo {author} {\bibfnamefont {R.}~\bibnamefont
  {Vasseur}}, \bibinfo {author} {\bibfnamefont {C.}~\bibnamefont {Karrasch}}, \
  and\ \bibinfo {author} {\bibfnamefont {J.~E.}\ \bibnamefont {Moore}},\ }\href
  {\doibase 10.1103/physrevb.97.045407} {\bibfield  {journal} {\bibinfo
  {journal} {Physical Review B}\ }\textbf {\bibinfo {volume} {97}} (\bibinfo
  {year} {2018}),\ 10.1103/physrevb.97.045407}\BibitemShut {NoStop}%
\bibitem [{\citenamefont {Doyon}\ and\ \citenamefont
  {Spohn}(2017)}]{SciPostPhys.3.6.039}%
  \BibitemOpen
  \bibfield  {author} {\bibinfo {author} {\bibfnamefont {B.}~\bibnamefont
  {Doyon}}\ and\ \bibinfo {author} {\bibfnamefont {H.}~\bibnamefont {Spohn}},\
  }\href {\doibase 10.21468/SciPostPhys.3.6.039} {\bibfield  {journal}
  {\bibinfo  {journal} {SciPost Phys.}\ }\textbf {\bibinfo {volume} {3}},\
  \bibinfo {pages} {039} (\bibinfo {year} {2017})}\BibitemShut {NoStop}%
\bibitem [{\citenamefont {Ilievski}\ and\ \citenamefont {{De
  Nardis}}(2017{\natexlab{b}})}]{IN_Hubbard}%
  \BibitemOpen
  \bibfield  {author} {\bibinfo {author} {\bibfnamefont {E.}~\bibnamefont
  {Ilievski}}\ and\ \bibinfo {author} {\bibfnamefont {J.}~\bibnamefont {{De
  Nardis}}},\ }\href {\doibase 10.1103/physrevb.96.081118} {\bibfield
  {journal} {\bibinfo  {journal} {Physical Review B}\ }\textbf {\bibinfo
  {volume} {96}} (\bibinfo {year} {2017}{\natexlab{b}}),\
  10.1103/physrevb.96.081118}\BibitemShut {NoStop}%
\bibitem [{\citenamefont {{De Nardis}}\ \emph {et~al.}(2018)\citenamefont {{De
  Nardis}}, \citenamefont {Bernard},\ and\ \citenamefont
  {Doyon}}]{DeNardis2018}%
  \BibitemOpen
  \bibfield  {author} {\bibinfo {author} {\bibfnamefont {J.}~\bibnamefont {{De
  Nardis}}}, \bibinfo {author} {\bibfnamefont {D.}~\bibnamefont {Bernard}}, \
  and\ \bibinfo {author} {\bibfnamefont {B.}~\bibnamefont {Doyon}},\ }\href
  {\doibase 10.1103/physrevlett.121.160603} {\bibfield  {journal} {\bibinfo
  {journal} {Physical Review Letters}\ }\textbf {\bibinfo {volume} {121}}
  (\bibinfo {year} {2018}),\ 10.1103/physrevlett.121.160603}\BibitemShut
  {NoStop}%
\bibitem [{\citenamefont {{De Nardis}}\ \emph
  {et~al.}(2019{\natexlab{a}})\citenamefont {{De Nardis}}, \citenamefont
  {Bernard},\ and\ \citenamefont {Doyon}}]{DeNardis_SciPost}%
  \BibitemOpen
  \bibfield  {author} {\bibinfo {author} {\bibfnamefont {J.}~\bibnamefont {{De
  Nardis}}}, \bibinfo {author} {\bibfnamefont {D.}~\bibnamefont {Bernard}}, \
  and\ \bibinfo {author} {\bibfnamefont {B.}~\bibnamefont {Doyon}},\ }\href
  {\doibase 10.21468/scipostphys.6.4.049} {\bibfield  {journal} {\bibinfo
  {journal} {{SciPost} Physics}\ }\textbf {\bibinfo {volume} {6}} (\bibinfo
  {year} {2019}{\natexlab{a}}),\ 10.21468/scipostphys.6.4.049}\BibitemShut
  {NoStop}%
\bibitem [{\citenamefont {Medenjak}\ \emph {et~al.}(2019)\citenamefont
  {Medenjak}, \citenamefont {De~Nardis},\ and\ \citenamefont
  {Yoshimura}}]{medenjak2019diffusion}%
  \BibitemOpen
  \bibfield  {author} {\bibinfo {author} {\bibfnamefont {M.}~\bibnamefont
  {Medenjak}}, \bibinfo {author} {\bibfnamefont {J.}~\bibnamefont {De~Nardis}},
  \ and\ \bibinfo {author} {\bibfnamefont {T.}~\bibnamefont {Yoshimura}},\
  }\href@noop {} {\  (\bibinfo {year} {2019})},\ \Eprint
  {http://arxiv.org/abs/arXiv:1911.01995} {arXiv:1911.01995} \BibitemShut
  {NoStop}%
\bibitem [{\citenamefont {Doyon}(2019)}]{doyon2019diffusion}%
  \BibitemOpen
  \bibfield  {author} {\bibinfo {author} {\bibfnamefont {B.}~\bibnamefont
  {Doyon}},\ }\href@noop {} {\  (\bibinfo {year} {2019})},\ \Eprint
  {http://arxiv.org/abs/arXiv:1912.01551} {arXiv:1912.01551} \BibitemShut
  {NoStop}%
\bibitem [{\citenamefont {Bulchandani}\ \emph {et~al.}(2017)\citenamefont
  {Bulchandani}, \citenamefont {Vasseur}, \citenamefont {Karrasch},\ and\
  \citenamefont {Moore}}]{Bulchandani2017}%
  \BibitemOpen
  \bibfield  {author} {\bibinfo {author} {\bibfnamefont {V.~B.}\ \bibnamefont
  {Bulchandani}}, \bibinfo {author} {\bibfnamefont {R.}~\bibnamefont
  {Vasseur}}, \bibinfo {author} {\bibfnamefont {C.}~\bibnamefont {Karrasch}}, \
  and\ \bibinfo {author} {\bibfnamefont {J.~E.}\ \bibnamefont {Moore}},\ }\href
  {\doibase 10.1103/physrevlett.119.220604} {\bibfield  {journal} {\bibinfo
  {journal} {Physical Review Letters}\ }\textbf {\bibinfo {volume} {119}}
  (\bibinfo {year} {2017}),\ 10.1103/physrevlett.119.220604}\BibitemShut
  {NoStop}%
\bibitem [{\citenamefont {Doyon}\ \emph {et~al.}(2018)\citenamefont {Doyon},
  \citenamefont {Yoshimura},\ and\ \citenamefont
  {Caux}}]{PhysRevLett.120.045301}%
  \BibitemOpen
  \bibfield  {author} {\bibinfo {author} {\bibfnamefont {B.}~\bibnamefont
  {Doyon}}, \bibinfo {author} {\bibfnamefont {T.}~\bibnamefont {Yoshimura}}, \
  and\ \bibinfo {author} {\bibfnamefont {J.-S.}\ \bibnamefont {Caux}},\ }\href
  {\doibase 10.1103/PhysRevLett.120.045301} {\bibfield  {journal} {\bibinfo
  {journal} {Phys. Rev. Lett.}\ }\textbf {\bibinfo {volume} {120}},\ \bibinfo
  {pages} {045301} (\bibinfo {year} {2018})}\BibitemShut {NoStop}%
\bibitem [{\citenamefont {Doyon}\ \emph {et~al.}(2017)\citenamefont {Doyon},
  \citenamefont {Dubail}, \citenamefont {Konik},\ and\ \citenamefont
  {Yoshimura}}]{PhysRevLett.119.195301}%
  \BibitemOpen
  \bibfield  {author} {\bibinfo {author} {\bibfnamefont {B.}~\bibnamefont
  {Doyon}}, \bibinfo {author} {\bibfnamefont {J.}~\bibnamefont {Dubail}},
  \bibinfo {author} {\bibfnamefont {R.}~\bibnamefont {Konik}}, \ and\ \bibinfo
  {author} {\bibfnamefont {T.}~\bibnamefont {Yoshimura}},\ }\href {\doibase
  10.1103/PhysRevLett.119.195301} {\bibfield  {journal} {\bibinfo  {journal}
  {Phys. Rev. Lett.}\ }\textbf {\bibinfo {volume} {119}},\ \bibinfo {pages}
  {195301} (\bibinfo {year} {2017})}\BibitemShut {NoStop}%
\bibitem [{\citenamefont {Schemmer}\ \emph {et~al.}(2018)\citenamefont
  {Schemmer}, \citenamefont {Bouchoule}, \citenamefont {Doyon},\ and\
  \citenamefont {Dubail}}]{1810.07170}%
  \BibitemOpen
  \bibfield  {author} {\bibinfo {author} {\bibfnamefont {M.}~\bibnamefont
  {Schemmer}}, \bibinfo {author} {\bibfnamefont {I.}~\bibnamefont {Bouchoule}},
  \bibinfo {author} {\bibfnamefont {B.}~\bibnamefont {Doyon}}, \ and\ \bibinfo
  {author} {\bibfnamefont {J.}~\bibnamefont {Dubail}},\ }\href@noop {} {\
  (\bibinfo {year} {2018})},\ \Eprint {http://arxiv.org/abs/arXiv:1810.07170}
  {arXiv:1810.07170} \BibitemShut {NoStop}%
\bibitem [{\citenamefont {Misguich}\ \emph {et~al.}(2017)\citenamefont
  {Misguich}, \citenamefont {Mallick},\ and\ \citenamefont
  {Krapivsky}}]{Misguich2017}%
  \BibitemOpen
  \bibfield  {author} {\bibinfo {author} {\bibfnamefont {G.}~\bibnamefont
  {Misguich}}, \bibinfo {author} {\bibfnamefont {K.}~\bibnamefont {Mallick}}, \
  and\ \bibinfo {author} {\bibfnamefont {P.~L.}\ \bibnamefont {Krapivsky}},\
  }\href {\doibase 10.1103/PhysRevB.96.195151} {\bibfield  {journal} {\bibinfo
  {journal} {Phys. Rev. B}\ }\textbf {\bibinfo {volume} {96}},\ \bibinfo
  {pages} {195151} (\bibinfo {year} {2017})}\BibitemShut {NoStop}%
\bibitem [{\citenamefont {Misguich}\ \emph {et~al.}(2019)\citenamefont
  {Misguich}, \citenamefont {Pavloff},\ and\ \citenamefont
  {Pasquier}}]{Misguich2019}%
  \BibitemOpen
  \bibfield  {author} {\bibinfo {author} {\bibfnamefont {G.}~\bibnamefont
  {Misguich}}, \bibinfo {author} {\bibfnamefont {N.}~\bibnamefont {Pavloff}}, \
  and\ \bibinfo {author} {\bibfnamefont {V.}~\bibnamefont {Pasquier}},\ }\href
  {\doibase 10.21468/SciPostPhys.7.2.025} {\bibfield  {journal} {\bibinfo
  {journal} {SciPost Phys.}\ }\textbf {\bibinfo {volume} {7}},\ \bibinfo
  {pages} {25} (\bibinfo {year} {2019})}\BibitemShut {NoStop}%
\bibitem [{\citenamefont {Gamayun}\ \emph {et~al.}(2019)\citenamefont
  {Gamayun}, \citenamefont {Miao},\ and\ \citenamefont
  {Ilievski}}]{Gamayun2019}%
  \BibitemOpen
  \bibfield  {author} {\bibinfo {author} {\bibfnamefont {O.}~\bibnamefont
  {Gamayun}}, \bibinfo {author} {\bibfnamefont {Y.}~\bibnamefont {Miao}}, \
  and\ \bibinfo {author} {\bibfnamefont {E.}~\bibnamefont {Ilievski}},\ }\href
  {\doibase 10.1103/physrevb.99.140301} {\bibfield  {journal} {\bibinfo
  {journal} {Physical Review B}\ }\textbf {\bibinfo {volume} {99}} (\bibinfo
  {year} {2019}),\ 10.1103/physrevb.99.140301}\BibitemShut {NoStop}%
\bibitem [{\citenamefont {{J. Sirker and R. G. Pereira and I.
  Affleck}}(2011)}]{Sirker2011}%
  \BibitemOpen
  \bibfield  {author} {\bibinfo {author} {\bibnamefont {{J. Sirker and R. G.
  Pereira and I. Affleck}}},\ }\href {\doibase 10.1103/physrevb.83.035115}
  {\bibfield  {journal} {\bibinfo  {journal} {Physical Review B}\ }\textbf
  {\bibinfo {volume} {83}} (\bibinfo {year} {2011}),\
  10.1103/physrevb.83.035115}\BibitemShut {NoStop}%
\bibitem [{\citenamefont {{C. Karrasch and J. Hauschild and S. Langer and F.
  Heidrich-Meisner}}(2013)}]{Karrasch2013}%
  \BibitemOpen
  \bibfield  {author} {\bibinfo {author} {\bibnamefont {{C. Karrasch and J.
  Hauschild and S. Langer and F. Heidrich-Meisner}}},\ }\href {\doibase
  10.1103/physrevb.87.245128} {\bibfield  {journal} {\bibinfo  {journal}
  {Physical Review B}\ }\textbf {\bibinfo {volume} {87}} (\bibinfo {year}
  {2013}),\ 10.1103/physrevb.87.245128}\BibitemShut {NoStop}%
\bibitem [{\citenamefont {Karrasch}\ \emph {et~al.}(2014)\citenamefont
  {Karrasch}, \citenamefont {Moore},\ and\ \citenamefont
  {Heidrich-Meisner}}]{Karrasch2014}%
  \BibitemOpen
  \bibfield  {author} {\bibinfo {author} {\bibfnamefont {C.}~\bibnamefont
  {Karrasch}}, \bibinfo {author} {\bibfnamefont {J.~E.}\ \bibnamefont {Moore}},
  \ and\ \bibinfo {author} {\bibfnamefont {F.}~\bibnamefont
  {Heidrich-Meisner}},\ }\href {\doibase 10.1103/physrevb.89.075139} {\bibfield
   {journal} {\bibinfo  {journal} {Physical Review B}\ }\textbf {\bibinfo
  {volume} {89}} (\bibinfo {year} {2014}),\
  10.1103/physrevb.89.075139}\BibitemShut {NoStop}%
\bibitem [{\citenamefont {Steinigeweg}\ \emph {et~al.}(2015)\citenamefont
  {Steinigeweg}, \citenamefont {Gemmer},\ and\ \citenamefont
  {Brenig}}]{Steinigeweg2015}%
  \BibitemOpen
  \bibfield  {author} {\bibinfo {author} {\bibfnamefont {R.}~\bibnamefont
  {Steinigeweg}}, \bibinfo {author} {\bibfnamefont {J.}~\bibnamefont {Gemmer}},
  \ and\ \bibinfo {author} {\bibfnamefont {W.}~\bibnamefont {Brenig}},\ }\href
  {\doibase 10.1103/PhysRevB.91.104404} {\bibfield  {journal} {\bibinfo
  {journal} {Phys. Rev. B}\ }\textbf {\bibinfo {volume} {91}},\ \bibinfo
  {pages} {104404} (\bibinfo {year} {2015})}\BibitemShut {NoStop}%
\bibitem [{\citenamefont {Steinigeweg}\ \emph {et~al.}(2017)\citenamefont
  {Steinigeweg}, \citenamefont {Jin}, \citenamefont {Schmidtke}, \citenamefont
  {De~Raedt}, \citenamefont {Michielsen},\ and\ \citenamefont
  {Gemmer}}]{Steinigeweg2017}%
  \BibitemOpen
  \bibfield  {author} {\bibinfo {author} {\bibfnamefont {R.}~\bibnamefont
  {Steinigeweg}}, \bibinfo {author} {\bibfnamefont {F.}~\bibnamefont {Jin}},
  \bibinfo {author} {\bibfnamefont {D.}~\bibnamefont {Schmidtke}}, \bibinfo
  {author} {\bibfnamefont {H.}~\bibnamefont {De~Raedt}}, \bibinfo {author}
  {\bibfnamefont {K.}~\bibnamefont {Michielsen}}, \ and\ \bibinfo {author}
  {\bibfnamefont {J.}~\bibnamefont {Gemmer}},\ }\href {\doibase
  10.1103/PhysRevB.95.035155} {\bibfield  {journal} {\bibinfo  {journal} {Phys.
  Rev. B}\ }\textbf {\bibinfo {volume} {95}},\ \bibinfo {pages} {035155}
  (\bibinfo {year} {2017})}\BibitemShut {NoStop}%
\bibitem [{\citenamefont {{\v{Z}}nidaric}(2011)}]{PhysRevLett.106.220601}%
  \BibitemOpen
  \bibfield  {author} {\bibinfo {author} {\bibfnamefont {M.}~\bibnamefont
  {{\v{Z}}nidaric}},\ }\href {\doibase 10.1103/PhysRevLett.106.220601}
  {\bibfield  {journal} {\bibinfo  {journal} {Phys. Rev. Lett.}\ }\textbf
  {\bibinfo {volume} {106}},\ \bibinfo {pages} {220601} (\bibinfo {year}
  {2011})}\BibitemShut {NoStop}%
\bibitem [{\citenamefont {Ljubotina}\ \emph {et~al.}(2017)\citenamefont
  {Ljubotina}, \citenamefont {{\v{Z}}nidari{\v{c}}},\ and\ \citenamefont
  {Prosen}}]{Ljubotina2017}%
  \BibitemOpen
  \bibfield  {author} {\bibinfo {author} {\bibfnamefont {M.}~\bibnamefont
  {Ljubotina}}, \bibinfo {author} {\bibfnamefont {M.}~\bibnamefont
  {{\v{Z}}nidari{\v{c}}}}, \ and\ \bibinfo {author} {\bibfnamefont
  {T.}~\bibnamefont {Prosen}},\ }\href {\doibase 10.1038/ncomms16117}
  {\bibfield  {journal} {\bibinfo  {journal} {Nature Communications}\ }\textbf
  {\bibinfo {volume} {8}} (\bibinfo {year} {2017}),\
  10.1038/ncomms16117}\BibitemShut {NoStop}%
\bibitem [{\citenamefont {Ilievski}\ \emph {et~al.}(2018)\citenamefont
  {Ilievski}, \citenamefont {{De Nardis}}, \citenamefont {Medenjak},\ and\
  \citenamefont {Prosen}}]{Ilievski2018}%
  \BibitemOpen
  \bibfield  {author} {\bibinfo {author} {\bibfnamefont {E.}~\bibnamefont
  {Ilievski}}, \bibinfo {author} {\bibfnamefont {J.}~\bibnamefont {{De
  Nardis}}}, \bibinfo {author} {\bibfnamefont {M.}~\bibnamefont {Medenjak}}, \
  and\ \bibinfo {author} {\bibfnamefont {T.}~\bibnamefont {Prosen}},\ }\href
  {\doibase 10.1103/physrevlett.121.230602} {\bibfield  {journal} {\bibinfo
  {journal} {Physical Review Letters}\ }\textbf {\bibinfo {volume} {121}}
  (\bibinfo {year} {2018}),\ 10.1103/physrevlett.121.230602}\BibitemShut
  {NoStop}%
\bibitem [{\citenamefont {Gopalakrishnan}\ and\ \citenamefont
  {Vasseur}(2019)}]{GV2019}%
  \BibitemOpen
  \bibfield  {author} {\bibinfo {author} {\bibfnamefont {S.}~\bibnamefont
  {Gopalakrishnan}}\ and\ \bibinfo {author} {\bibfnamefont {R.}~\bibnamefont
  {Vasseur}},\ }\href {\doibase 10.1103/physrevlett.122.127202} {\bibfield
  {journal} {\bibinfo  {journal} {Physical Review Letters}\ }\textbf {\bibinfo
  {volume} {122}} (\bibinfo {year} {2019}),\
  10.1103/physrevlett.122.127202}\BibitemShut {NoStop}%
\bibitem [{\citenamefont {{De Nardis}}\ \emph
  {et~al.}(2019{\natexlab{b}})\citenamefont {{De Nardis}}, \citenamefont
  {Medenjak}, \citenamefont {Karrasch},\ and\ \citenamefont
  {Ilievski}}]{NMKI2019}%
  \BibitemOpen
  \bibfield  {author} {\bibinfo {author} {\bibfnamefont {J.}~\bibnamefont {{De
  Nardis}}}, \bibinfo {author} {\bibfnamefont {M.}~\bibnamefont {Medenjak}},
  \bibinfo {author} {\bibfnamefont {C.}~\bibnamefont {Karrasch}}, \ and\
  \bibinfo {author} {\bibfnamefont {E.}~\bibnamefont {Ilievski}},\ }\href
  {\doibase 10.1103/physrevlett.123.186601} {\bibfield  {journal} {\bibinfo
  {journal} {Physical Review Letters}\ }\textbf {\bibinfo {volume} {123}}
  (\bibinfo {year} {2019}{\natexlab{b}}),\
  10.1103/physrevlett.123.186601}\BibitemShut {NoStop}%
\bibitem [{\citenamefont {Dupont}\ and\ \citenamefont
  {Moore}(2019)}]{DupontMoore2019}%
  \BibitemOpen
  \bibfield  {author} {\bibinfo {author} {\bibfnamefont {M.}~\bibnamefont
  {Dupont}}\ and\ \bibinfo {author} {\bibfnamefont {J.~E.}\ \bibnamefont
  {Moore}},\ }\href@noop {} {\  (\bibinfo {year} {2019})},\ \Eprint
  {http://arxiv.org/abs/arXiv:1907.12115} {arXiv:1907.12115} \BibitemShut
  {NoStop}%
\bibitem [{\citenamefont {Bulchandani}(2019)}]{Vir2019}%
  \BibitemOpen
  \bibfield  {author} {\bibinfo {author} {\bibfnamefont {V.~B.}\ \bibnamefont
  {Bulchandani}},\ }\href@noop {} {\  (\bibinfo {year} {2019})},\ \Eprint
  {http://arxiv.org/abs/arXiv:1910.08266} {arXiv:1910.08266} \BibitemShut
  {NoStop}%
\bibitem [{\citenamefont {Ljubotina}\ \emph {et~al.}(2019)\citenamefont
  {Ljubotina}, \citenamefont {{\v{Z}}nidari{\v{c}}},\ and\ \citenamefont
  {Prosen}}]{Ljubotina2019}%
  \BibitemOpen
  \bibfield  {author} {\bibinfo {author} {\bibfnamefont {M.}~\bibnamefont
  {Ljubotina}}, \bibinfo {author} {\bibfnamefont {M.}~\bibnamefont
  {{\v{Z}}nidari{\v{c}}}}, \ and\ \bibinfo {author} {\bibfnamefont
  {T.}~\bibnamefont {Prosen}},\ }\href {\doibase
  10.1103/physrevlett.122.210602} {\bibfield  {journal} {\bibinfo  {journal}
  {Physical Review Letters}\ }\textbf {\bibinfo {volume} {122}} (\bibinfo
  {year} {2019}),\ 10.1103/physrevlett.122.210602}\BibitemShut {NoStop}%
\bibitem [{\citenamefont {Das}\ \emph {et~al.}(2019)\citenamefont {Das},
  \citenamefont {Kulkarni}, \citenamefont {Spohn},\ and\ \citenamefont
  {Dhar}}]{PhysRevE.100.042116}%
  \BibitemOpen
  \bibfield  {author} {\bibinfo {author} {\bibfnamefont {A.}~\bibnamefont
  {Das}}, \bibinfo {author} {\bibfnamefont {M.}~\bibnamefont {Kulkarni}},
  \bibinfo {author} {\bibfnamefont {H.}~\bibnamefont {Spohn}}, \ and\ \bibinfo
  {author} {\bibfnamefont {A.}~\bibnamefont {Dhar}},\ }\href {\doibase
  10.1103/PhysRevE.100.042116} {\bibfield  {journal} {\bibinfo  {journal}
  {Phys. Rev. E}\ }\textbf {\bibinfo {volume} {100}},\ \bibinfo {pages}
  {042116} (\bibinfo {year} {2019})}\BibitemShut {NoStop}%
\bibitem [{\citenamefont {\v{Z}. Krajnik}\ and\ \citenamefont
  {Prosen}(2019)}]{Krajnik2019}%
  \BibitemOpen
  \bibfield  {author} {\bibinfo {author} {\bibnamefont {\v{Z}. Krajnik}}\ and\
  \bibinfo {author} {\bibfnamefont {T.}~\bibnamefont {Prosen}},\ }\href@noop {}
  {\  (\bibinfo {year} {2019})},\ \Eprint
  {http://arxiv.org/abs/arXiv:1909.03799} {arXiv:1909.03799} \BibitemShut
  {NoStop}%
\bibitem [{\citenamefont {Bagchi}(2013)}]{Bagchi2013}%
  \BibitemOpen
  \bibfield  {author} {\bibinfo {author} {\bibfnamefont {D.}~\bibnamefont
  {Bagchi}},\ }\href {\doibase 10.1103/physrevb.87.075133} {\bibfield
  {journal} {\bibinfo  {journal} {Physical Review B}\ }\textbf {\bibinfo
  {volume} {87}} (\bibinfo {year} {2013}),\
  10.1103/physrevb.87.075133}\BibitemShut {NoStop}%
\bibitem [{\citenamefont {{Gerhard M\"{u}ller}}(1988)}]{Muller1988}%
  \BibitemOpen
  \bibfield  {author} {\bibinfo {author} {\bibnamefont {{Gerhard
  M\"{u}ller}}},\ }\href {\doibase 10.1103/physrevlett.60.2785} {\bibfield
  {journal} {\bibinfo  {journal} {Physical Review Letters}\ }\textbf {\bibinfo
  {volume} {60}},\ \bibinfo {pages} {2785} (\bibinfo {year}
  {1988})}\BibitemShut {NoStop}%
\bibitem [{\citenamefont {Gerling}\ and\ \citenamefont
  {Landau}(1990)}]{Gerling1990}%
  \BibitemOpen
  \bibfield  {author} {\bibinfo {author} {\bibfnamefont {R.~W.}\ \bibnamefont
  {Gerling}}\ and\ \bibinfo {author} {\bibfnamefont {D.~P.}\ \bibnamefont
  {Landau}},\ }\href {\doibase 10.1103/physrevb.42.8214} {\bibfield  {journal}
  {\bibinfo  {journal} {Physical Review B}\ }\textbf {\bibinfo {volume} {42}},\
  \bibinfo {pages} {8214} (\bibinfo {year} {1990})}\BibitemShut {NoStop}%
\bibitem [{\citenamefont {{Jian-Min Liu and Niraj Srivastava and V. S.
  Viswanath and Gerhard M\"{u}ller}}(1991)}]{Liu1991}%
  \BibitemOpen
  \bibfield  {author} {\bibinfo {author} {\bibnamefont {{Jian-Min Liu and Niraj
  Srivastava and V. S. Viswanath and Gerhard M\"{u}ller}}},\ }\href {\doibase
  10.1063/1.350037} {\bibfield  {journal} {\bibinfo  {journal} {Journal of
  Applied Physics}\ }\textbf {\bibinfo {volume} {70}},\ \bibinfo {pages} {6181}
  (\bibinfo {year} {1991})}\BibitemShut {NoStop}%
\bibitem [{\citenamefont {de~Alcantara~Bonfim}\ and\ \citenamefont
  {Reiter}(1992)}]{Bonfim1992}%
  \BibitemOpen
  \bibfield  {author} {\bibinfo {author} {\bibfnamefont {O.~F.}\ \bibnamefont
  {de~Alcantara~Bonfim}}\ and\ \bibinfo {author} {\bibfnamefont
  {G.}~\bibnamefont {Reiter}},\ }\href {\doibase 10.1103/physrevlett.69.367}
  {\bibfield  {journal} {\bibinfo  {journal} {Physical Review Letters}\
  }\textbf {\bibinfo {volume} {69}},\ \bibinfo {pages} {367} (\bibinfo {year}
  {1992})}\BibitemShut {NoStop}%
\bibitem [{\citenamefont {{Markus B\"{o}hm and Rainer W. Gerling and Hajo
  Leschke}}(1993)}]{Bohm1993}%
  \BibitemOpen
  \bibfield  {author} {\bibinfo {author} {\bibnamefont {{Markus B\"{o}hm and
  Rainer W. Gerling and Hajo Leschke}}},\ }\href {\doibase
  10.1103/physrevlett.70.248} {\bibfield  {journal} {\bibinfo  {journal}
  {Physical Review Letters}\ }\textbf {\bibinfo {volume} {70}},\ \bibinfo
  {pages} {248} (\bibinfo {year} {1993})}\BibitemShut {NoStop}%
\bibitem [{\citenamefont {{Niraj Srivastava and Jian-Min Liu and V. S.
  Viswanath and Gerhard M\"{u}ller}}(1994)}]{Srivastava1994}%
  \BibitemOpen
  \bibfield  {author} {\bibinfo {author} {\bibnamefont {{Niraj Srivastava and
  Jian-Min Liu and V. S. Viswanath and Gerhard M\"{u}ller}}},\ }\href {\doibase
  10.1063/1.356842} {\bibfield  {journal} {\bibinfo  {journal} {Journal of
  Applied Physics}\ }\textbf {\bibinfo {volume} {75}},\ \bibinfo {pages} {6751}
  (\bibinfo {year} {1994})}\BibitemShut {NoStop}%
\bibitem [{Note1()}]{Note1}%
  \BibitemOpen
  \bibinfo {note} {For spin-$S$ quantum chains linear in spin generators this
  is achieved by taking the expectation value on the $SU(2)$ spin-coherent
  states. Spin chains with nonlinear terms, where one encounters path-integral
  anomalies \cite {Wilson2011}, have to be excluded from our
  framework.}\BibitemShut {Stop}%
\bibitem [{\citenamefont {Herbrych}\ \emph {et~al.}(2011)\citenamefont
  {Herbrych}, \citenamefont {Prelov\ifmmode~\check{s}\else \v{s}\fi{}ek},\ and\
  \citenamefont {Zotos}}]{PhysRevB.84.155125}%
  \BibitemOpen
  \bibfield  {author} {\bibinfo {author} {\bibfnamefont {J.}~\bibnamefont
  {Herbrych}}, \bibinfo {author} {\bibfnamefont {P.}~\bibnamefont
  {Prelov\ifmmode~\check{s}\else \v{s}\fi{}ek}}, \ and\ \bibinfo {author}
  {\bibfnamefont {X.}~\bibnamefont {Zotos}},\ }\href {\doibase
  10.1103/PhysRevB.84.155125} {\bibfield  {journal} {\bibinfo  {journal} {Phys.
  Rev. B}\ }\textbf {\bibinfo {volume} {84}},\ \bibinfo {pages} {155125}
  (\bibinfo {year} {2011})}\BibitemShut {NoStop}%
\bibitem [{Note2()}]{Note2}%
  \BibitemOpen
  \bibinfo {note} {While in the {Landau--Lifshitz} hierarchy of commuting flows
  helices are stationary states of the full nonlinear Hamiltonian dynamics,
  their fate in generic spin systems is less clear. We nonetheless expect they
  become stabilized by effective decoupling of the curvature field at large
  times.}\BibitemShut {Stop}%
\bibitem [{\citenamefont {Das}\ \emph {et~al.}(2018{\natexlab{b}})\citenamefont
  {Das}, \citenamefont {Damle}, \citenamefont {Dhar}, \citenamefont {Huse},
  \citenamefont {Kulkarni}, \citenamefont {Mendl},\ and\ \citenamefont
  {Spohn}}]{Das2019}%
  \BibitemOpen
  \bibfield  {author} {\bibinfo {author} {\bibfnamefont {A.}~\bibnamefont
  {Das}}, \bibinfo {author} {\bibfnamefont {K.}~\bibnamefont {Damle}}, \bibinfo
  {author} {\bibfnamefont {A.}~\bibnamefont {Dhar}}, \bibinfo {author}
  {\bibfnamefont {D.~A.}\ \bibnamefont {Huse}}, \bibinfo {author}
  {\bibfnamefont {M.}~\bibnamefont {Kulkarni}}, \bibinfo {author}
  {\bibfnamefont {C.~B.}\ \bibnamefont {Mendl}}, \ and\ \bibinfo {author}
  {\bibfnamefont {H.}~\bibnamefont {Spohn}},\ }\href {\doibase
  10.1007/s10955-019-02397-y} {\  (\bibinfo {year} {2018}{\natexlab{b}}),\
  10.1007/s10955-019-02397-y},\ \Eprint {http://arxiv.org/abs/arXiv:1901.00024}
  {arXiv:1901.00024} \BibitemShut {NoStop}%
\bibitem [{\citenamefont {Barraquand}\ \emph {et~al.}(2019)\citenamefont
  {Barraquand}, \citenamefont {Doussal},\ and\ \citenamefont
  {Rosso}}]{1909.11557}%
  \BibitemOpen
  \bibfield  {author} {\bibinfo {author} {\bibfnamefont {G.}~\bibnamefont
  {Barraquand}}, \bibinfo {author} {\bibfnamefont {P.~L.}\ \bibnamefont
  {Doussal}}, \ and\ \bibinfo {author} {\bibfnamefont {A.}~\bibnamefont
  {Rosso}},\ }\href@noop {} {\  (\bibinfo {year} {2019})},\ \Eprint
  {http://arxiv.org/abs/arXiv:1909.11557} {arXiv:1909.11557} \BibitemShut
  {NoStop}%
\bibitem [{PB()}]{PB}%
  \BibitemOpen
  \href@noop {} {}\bibinfo {note} {P. Le Doussal, G. Barraquand, Private
  Comunication.}\BibitemShut {Stop}%
\bibitem [{SM()}]{SM}%
  \BibitemOpen
  \href@noop {} {}\bibinfo {note} {Supplemental Material associated with this
  manuscript.}\BibitemShut {Stop}%
\bibitem [{\citenamefont {Takhtajan}(1977)}]{Takhtajan1977}%
  \BibitemOpen
  \bibfield  {author} {\bibinfo {author} {\bibfnamefont {L.}~\bibnamefont
  {Takhtajan}},\ }\href {\doibase 10.1016/0375-9601(77)90727-7} {\bibfield
  {journal} {\bibinfo  {journal} {Physics Letters A}\ }\textbf {\bibinfo
  {volume} {64}},\ \bibinfo {pages} {235} (\bibinfo {year} {1977})}\BibitemShut
  {NoStop}%
\bibitem [{\citenamefont {Faddeev}\ and\ \citenamefont
  {Takhtajan}(1987)}]{Faddeev1987}%
  \BibitemOpen
  \bibfield  {author} {\bibinfo {author} {\bibfnamefont {L.~D.}\ \bibnamefont
  {Faddeev}}\ and\ \bibinfo {author} {\bibfnamefont {L.~A.}\ \bibnamefont
  {Takhtajan}},\ }\href {\doibase 10.1007/978-3-540-69969-9} {\emph {\bibinfo
  {title} {Hamiltonian Methods in the Theory of Solitons}}}\ (\bibinfo
  {publisher} {Springer Berlin Heidelberg},\ \bibinfo {year}
  {1987})\BibitemShut {NoStop}%
\bibitem [{\citenamefont {Derrida}\ \emph {et~al.}(1991)\citenamefont
  {Derrida}, \citenamefont {Lebowitz}, \citenamefont {Speer},\ and\
  \citenamefont {Spohn}}]{Derrida1991}%
  \BibitemOpen
  \bibfield  {author} {\bibinfo {author} {\bibfnamefont {B.}~\bibnamefont
  {Derrida}}, \bibinfo {author} {\bibfnamefont {J.~L.}\ \bibnamefont
  {Lebowitz}}, \bibinfo {author} {\bibfnamefont {E.~R.}\ \bibnamefont {Speer}},
  \ and\ \bibinfo {author} {\bibfnamefont {H.}~\bibnamefont {Spohn}},\ }\href
  {\doibase 10.1103/PhysRevLett.67.165} {\bibfield  {journal} {\bibinfo
  {journal} {Phys. Rev. Lett.}\ }\textbf {\bibinfo {volume} {67}},\ \bibinfo
  {pages} {165} (\bibinfo {year} {1991})}\BibitemShut {NoStop}%
\bibitem [{\citenamefont {Devillard}\ and\ \citenamefont
  {Spohn}(1992)}]{Devillard1992}%
  \BibitemOpen
  \bibfield  {author} {\bibinfo {author} {\bibfnamefont {P.}~\bibnamefont
  {Devillard}}\ and\ \bibinfo {author} {\bibfnamefont {H.}~\bibnamefont
  {Spohn}},\ }\href {\doibase 10.1007/bf01055718} {\bibfield  {journal}
  {\bibinfo  {journal} {Journal of Statistical Physics}\ }\textbf {\bibinfo
  {volume} {66}},\ \bibinfo {pages} {1089} (\bibinfo {year}
  {1992})}\BibitemShut {NoStop}%
\bibitem [{\citenamefont {Paczuski}\ \emph {et~al.}(1992)\citenamefont
  {Paczuski}, \citenamefont {Barma}, \citenamefont {Majumdar},\ and\
  \citenamefont {Hwa}}]{Paczuski1992}%
  \BibitemOpen
  \bibfield  {author} {\bibinfo {author} {\bibfnamefont {M.}~\bibnamefont
  {Paczuski}}, \bibinfo {author} {\bibfnamefont {M.}~\bibnamefont {Barma}},
  \bibinfo {author} {\bibfnamefont {S.~N.}\ \bibnamefont {Majumdar}}, \ and\
  \bibinfo {author} {\bibfnamefont {T.}~\bibnamefont {Hwa}},\ }\href {\doibase
  10.1103/PhysRevLett.69.2735} {\bibfield  {journal} {\bibinfo  {journal}
  {Phys. Rev. Lett.}\ }\textbf {\bibinfo {volume} {69}},\ \bibinfo {pages}
  {2735} (\bibinfo {year} {1992})}\BibitemShut {NoStop}%
\bibitem [{\citenamefont {Kazakov}\ \emph {et~al.}(2004)\citenamefont
  {Kazakov}, \citenamefont {Marshakov}, \citenamefont {A.Minahan},\ and\
  \citenamefont {Zarembo}}]{KMMZ2004}%
  \BibitemOpen
  \bibfield  {author} {\bibinfo {author} {\bibfnamefont {V.~A.}\ \bibnamefont
  {Kazakov}}, \bibinfo {author} {\bibfnamefont {A.}~\bibnamefont {Marshakov}},
  \bibinfo {author} {\bibfnamefont {J.}~\bibnamefont {A.Minahan}}, \ and\
  \bibinfo {author} {\bibfnamefont {K.}~\bibnamefont {Zarembo}},\ }\href
  {\doibase 10.1088/1126-6708/2004/05/024} {\bibfield  {journal} {\bibinfo
  {journal} {Journal of High Energy Physics}\ }\textbf {\bibinfo {volume}
  {2004}},\ \bibinfo {pages} {024} (\bibinfo {year} {2004})}\BibitemShut
  {NoStop}%
\bibitem [{\citenamefont {Bargheer}\ \emph {et~al.}(2008)\citenamefont
  {Bargheer}, \citenamefont {Beisert},\ and\ \citenamefont
  {Gromov}}]{Bargheer2008}%
  \BibitemOpen
  \bibfield  {author} {\bibinfo {author} {\bibfnamefont {T.}~\bibnamefont
  {Bargheer}}, \bibinfo {author} {\bibfnamefont {N.}~\bibnamefont {Beisert}}, \
  and\ \bibinfo {author} {\bibfnamefont {N.}~\bibnamefont {Gromov}},\ }\href
  {\doibase 10.1088/1367-2630/10/10/103023} {\bibfield  {journal} {\bibinfo
  {journal} {New Journal of Physics}\ }\textbf {\bibinfo {volume} {10}},\
  \bibinfo {pages} {103023} (\bibinfo {year} {2008})}\BibitemShut {NoStop}%
\bibitem [{\citenamefont {Li}(2019)}]{Li2019}%
  \BibitemOpen
  \bibfield  {author} {\bibinfo {author} {\bibfnamefont {N.}~\bibnamefont
  {Li}},\ }\href {\doibase 10.1103/PhysRevE.100.062104} {\bibfield  {journal}
  {\bibinfo  {journal} {Phys. Rev. E}\ }\textbf {\bibinfo {volume} {100}},\
  \bibinfo {pages} {062104} (\bibinfo {year} {2019})}\BibitemShut {NoStop}%
\bibitem [{\citenamefont {White}(1992)}]{white1992}%
  \BibitemOpen
  \bibfield  {author} {\bibinfo {author} {\bibfnamefont {S.~R.}\ \bibnamefont
  {White}},\ }\href {\doibase 10.1103/physrevlett.69.2863} {\bibfield
  {journal} {\bibinfo  {journal} {Physical Review Letters}\ }\textbf {\bibinfo
  {volume} {69}},\ \bibinfo {pages} {2863} (\bibinfo {year}
  {1992})}\BibitemShut {NoStop}%
\bibitem [{\citenamefont {{Ulrich Schollw\"{o}ck}}(2011)}]{Schollwoeck2011}%
  \BibitemOpen
  \bibfield  {author} {\bibinfo {author} {\bibnamefont {{Ulrich
  Schollw\"{o}ck}}},\ }\href {\doibase 10.1016/j.aop.2010.09.012} {\bibfield
  {journal} {\bibinfo  {journal} {Annals of Physics}\ }\textbf {\bibinfo
  {volume} {326}},\ \bibinfo {pages} {96} (\bibinfo {year} {2011})}\BibitemShut
  {NoStop}%
\bibitem [{\citenamefont {Karrasch}\ \emph {et~al.}(2012)\citenamefont
  {Karrasch}, \citenamefont {Bardarson},\ and\ \citenamefont
  {Moore}}]{Karrasch2012}%
  \BibitemOpen
  \bibfield  {author} {\bibinfo {author} {\bibfnamefont {C.}~\bibnamefont
  {Karrasch}}, \bibinfo {author} {\bibfnamefont {J.~H.}\ \bibnamefont
  {Bardarson}}, \ and\ \bibinfo {author} {\bibfnamefont {J.~E.}\ \bibnamefont
  {Moore}},\ }\href {\doibase 10.1103/physrevlett.108.227206} {\bibfield
  {journal} {\bibinfo  {journal} {Physical Review Letters}\ }\textbf {\bibinfo
  {volume} {108}} (\bibinfo {year} {2012}),\
  10.1103/physrevlett.108.227206}\BibitemShut {NoStop}%
\bibitem [{\citenamefont {Richter}\ \emph {et~al.}(2019)\citenamefont
  {Richter}, \citenamefont {Casper}, \citenamefont {Brenig},\ and\
  \citenamefont {Steinigeweg}}]{Richter2019}%
  \BibitemOpen
  \bibfield  {author} {\bibinfo {author} {\bibfnamefont {J.}~\bibnamefont
  {Richter}}, \bibinfo {author} {\bibfnamefont {N.}~\bibnamefont {Casper}},
  \bibinfo {author} {\bibfnamefont {W.}~\bibnamefont {Brenig}}, \ and\ \bibinfo
  {author} {\bibfnamefont {R.}~\bibnamefont {Steinigeweg}},\ }\href {\doibase
  10.1103/PhysRevB.100.144423} {\bibfield  {journal} {\bibinfo  {journal}
  {Phys. Rev. B}\ }\textbf {\bibinfo {volume} {100}},\ \bibinfo {pages}
  {144423} (\bibinfo {year} {2019})}\BibitemShut {NoStop}%
\bibitem [{\citenamefont {Huang}\ \emph {et~al.}(2013)\citenamefont {Huang},
  \citenamefont {Karrasch},\ and\ \citenamefont {Moore}}]{Huang2013}%
  \BibitemOpen
  \bibfield  {author} {\bibinfo {author} {\bibfnamefont {Y.}~\bibnamefont
  {Huang}}, \bibinfo {author} {\bibfnamefont {C.}~\bibnamefont {Karrasch}}, \
  and\ \bibinfo {author} {\bibfnamefont {J.~E.}\ \bibnamefont {Moore}},\ }\href
  {\doibase 10.1103/PhysRevB.88.115126} {\bibfield  {journal} {\bibinfo
  {journal} {Phys. Rev. B}\ }\textbf {\bibinfo {volume} {88}},\ \bibinfo
  {pages} {115126} (\bibinfo {year} {2013})}\BibitemShut {NoStop}%
\bibitem [{\citenamefont {Vasseur}\ and\ \citenamefont
  {Moore}(2016)}]{Vasseur_2016}%
  \BibitemOpen
  \bibfield  {author} {\bibinfo {author} {\bibfnamefont {R.}~\bibnamefont
  {Vasseur}}\ and\ \bibinfo {author} {\bibfnamefont {J.~E.}\ \bibnamefont
  {Moore}},\ }\href {\doibase 10.1088/1742-5468/2016/06/064010} {\bibfield
  {journal} {\bibinfo  {journal} {Journal of Statistical Mechanics: Theory and
  Experiment}\ }\textbf {\bibinfo {volume} {2016}},\ \bibinfo {pages} {064010}
  (\bibinfo {year} {2016})}\BibitemShut {NoStop}%
\bibitem [{\citenamefont {Wilson}\ and\ \citenamefont
  {Galitski}(2011)}]{Wilson2011}%
  \BibitemOpen
  \bibfield  {author} {\bibinfo {author} {\bibfnamefont {J.~H.}\ \bibnamefont
  {Wilson}}\ and\ \bibinfo {author} {\bibfnamefont {V.}~\bibnamefont
  {Galitski}},\ }\href {\doibase 10.1103/physrevlett.106.110401} {\bibfield
  {journal} {\bibinfo  {journal} {Physical Review Letters}\ }\textbf {\bibinfo
  {volume} {106}} (\bibinfo {year} {2011}),\
  10.1103/physrevlett.106.110401}\BibitemShut {NoStop}%
\bibitem [{\citenamefont {Da~Rios}(1906)}]{DaRios}%
  \BibitemOpen
  \bibfield  {author} {\bibinfo {author} {\bibfnamefont {L.}~\bibnamefont
  {Da~Rios}},\ }\href@noop {} {\bibfield  {journal} {\bibinfo  {journal} {Rend.
  Circ. Mat. Palermo}\ }\textbf {\bibinfo {volume} {22}},\ \bibinfo {pages}
  {117} (\bibinfo {year} {1906})}\BibitemShut {NoStop}%
\bibitem [{\citenamefont {Barros}\ \emph {et~al.}(1999)\citenamefont {Barros},
  \citenamefont {Ferr{\'{a}}ndez}, \citenamefont {Lucas},\ and\ \citenamefont
  {Mero{\~{n}}o}}]{Barros1999}%
  \BibitemOpen
  \bibfield  {author} {\bibinfo {author} {\bibfnamefont {M.}~\bibnamefont
  {Barros}}, \bibinfo {author} {\bibfnamefont {A.}~\bibnamefont
  {Ferr{\'{a}}ndez}}, \bibinfo {author} {\bibfnamefont {P.}~\bibnamefont
  {Lucas}}, \ and\ \bibinfo {author} {\bibfnamefont {M.~A.}\ \bibnamefont
  {Mero{\~{n}}o}},\ }\href {\doibase 10.1016/s0393-0440(99)00005-4} {\bibfield
  {journal} {\bibinfo  {journal} {Journal of Geometry and Physics}\ }\textbf
  {\bibinfo {volume} {31}},\ \bibinfo {pages} {217} (\bibinfo {year}
  {1999})}\BibitemShut {NoStop}%
\bibitem [{\citenamefont {Fuchssteiner}(1983)}]{Fuchssteiner1983}%
  \BibitemOpen
  \bibfield  {author} {\bibinfo {author} {\bibfnamefont {B.}~\bibnamefont
  {Fuchssteiner}},\ }\href {\doibase 10.1143/ptp.70.1508} {\bibfield  {journal}
  {\bibinfo  {journal} {Progress of Theoretical Physics}\ }\textbf {\bibinfo
  {volume} {70}},\ \bibinfo {pages} {1508} (\bibinfo {year}
  {1983})}\BibitemShut {NoStop}%
\end{thebibliography}%
\clearpage
\newpage
\appendix
\onecolumngrid

\begin{center}
\textbf{{\Large Supplemental Material}}
\end{center}
\begin{center}
\textbf{{\large Universality classes of spin transport in one-dimensional isotropic magnets:\\
the onset of logarithmic anomalies}}
\end{center}

\section{On the numerical simulation of classical spin dynamics}

\begin{figure}[htb]
\centering
\includegraphics[width=0.4\columnwidth]{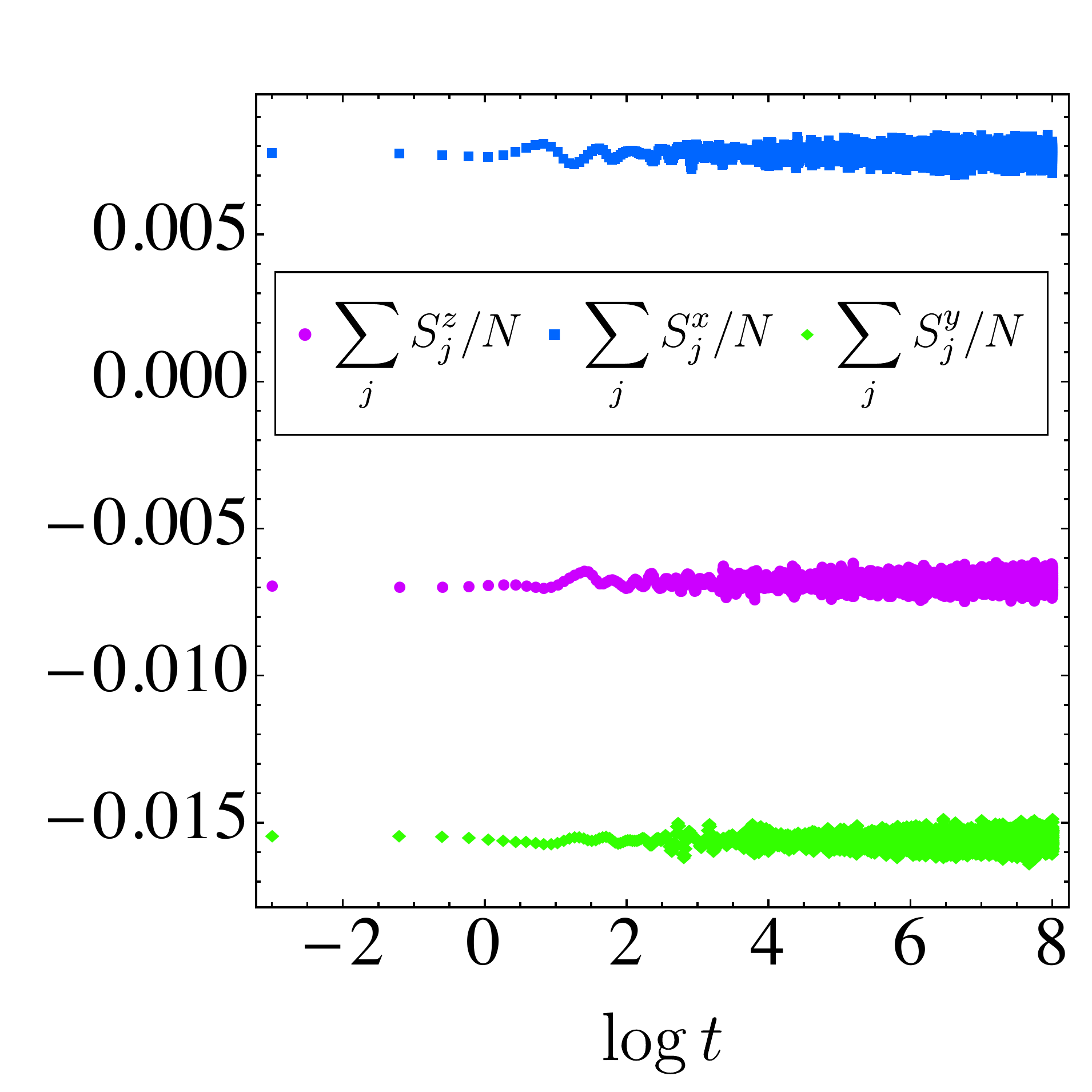}
\caption{Time-evolution of the approximately conserved magnetizations in the classical (non-integrable)
Heisenberg chain with integration step-size $\Delta t= 0.025$ and $L=1000$, starting from a single random initial state.}
\label{fig:error1}
\end{figure}

\begin{figure}[htb]
\centering
\includegraphics[width=0.4\columnwidth]{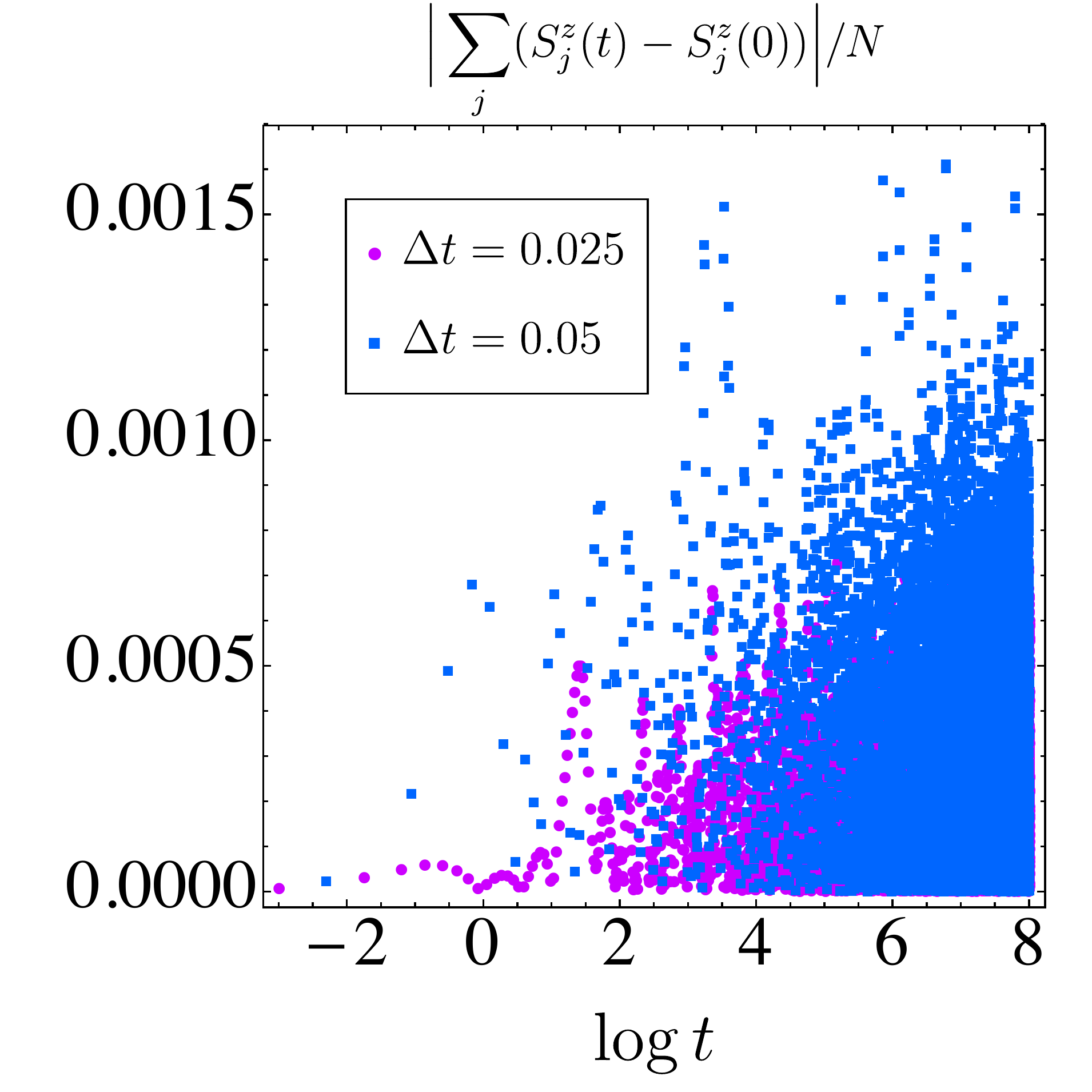}
\caption{Total magnetization fluctuations $|\sum_j S^z_j(t) -  \sum_j S^z_j(0)|/N$ as a function of time for the classical (non-integrable) Heisenberg chain for two different values of $\Delta t$ and $L=1000$, starting from a single random initial state.}
\label{fig:error2}
\end{figure}

\begin{figure}[htb]
\centering
\includegraphics[width=0.4\columnwidth]{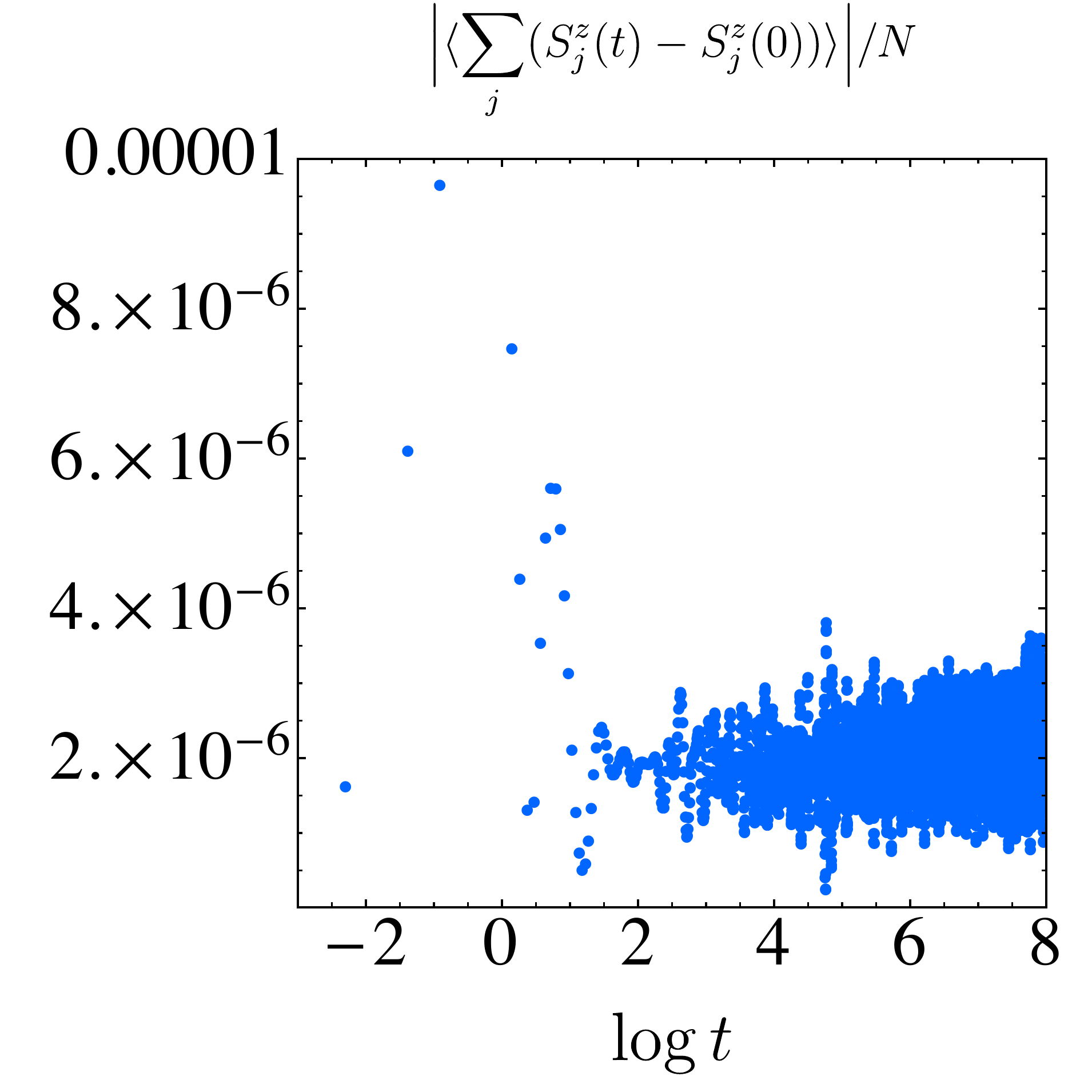}
\caption{Same as in Fig. \ref{fig:error2} with $\Delta t = 0.05$ but after averaging over $10^6$ random initial states in order to describe an infinite temperature ensemble.}
\label{fig:error3}
\end{figure}

We have carried out numerical simulations of classical spin dynamics using the method introduced in \cite{Bagchi2013}. The latter
proves to be superior compared to more standard Euler or Runge-Kutta integration schemes (as e.g. employed in \cite{Li2019}),
owing to \textit{exact} conservation of both total energy and spin magnitude $|\vec{S}_j|=1$ at each lattice site.
The method employs the fact that any equation of motion of type
\begin{equation}
\partial_t \vec{S}_j =  \vec{S}_j \times \vec{B}_j,
\end{equation}  
with vector $\vec{B}_j$ depending in general on the spins $\vec{S}_{j+1}, \vec{S}_{j-1} \ldots , \vec{S}_{j+n},\vec{S}_{j-n}$
can be analytically integrated with aid of the Rodrigues' rotational formula, yielding
\begin{equation}
\vec{S}_j (t + \Delta t) = [\vec{S} \cos \phi + \vec{S} \times \hat{B} \sin \phi + (\vec{S} \cdot \hat{B}) (1- \cos \phi) \hat{B}]_j(t)
\end{equation} 
where $\phi = |\vec{B}_i \Delta t|$ and $\hat{B} = \vec{B}/|\vec{B}|$. By evolving first $\vec{S}_j,\vec{S}_{j+ n}, \ldots \vec{S}_{j+ 2 n}$ and then $\vec{S}_{j+1}, \vec{S}_{j+ 1 + n},\ldots$ and so on, energy and $|\vec{S}_j|$ are conserved to all orders in
$\Delta t$, while all other local conserved quantities fluctuate within a range of order $(\Delta t)^3$ even on time-scales of
the order $t \sim 3000$, see Fig. \ref{fig:error1} and \ref{fig:error2}. In order to simulate dynamics at infinite temperature,
we have computed an average over $5 \times 10^5 - 10^6$ random initial states, which also reduces the error at large times, see Fig. \ref{fig:error3}. Moreover we stress that a large number of states is of fundamental importance in order to well recognise the logarithmic corrections at large times.  In order to reduce the noise, we have additionally performed
the ergodic time-average $t^{-1} \int_0^t \dd t' $ of the spin-spin correlations $C_j(t)$. \\

We stress that our numerical results do not differ from the ones reported in \cite{Bagchi2013}, where the presence of logarithmic corrections to diffusion was however missed in the analysis of the data. We instead believe that the numerical data at infinite temperature in \cite{Li2019} are incorrect, as they show normal diffusion, probably due to the Runge-Kutta integration schemes or lack of proper averaging on initial conditions.  The results in \cite{Li2019} for slightly anomalous diffusion at lower temperature are instead in agreement with the presence of logarithmic corrections. 

\section{Spin transport in the hierarchy of the quantum Heisenberg model}

We shall employ the toolbox of the generalized hydrodynamics \cite{PhysRevX.6.041065,PhysRevLett.117.207201} to examine the spin 
diffusion constant in the quantum Heisenberg hierarchy.
For definiteness we shall specialize here to the fundamental spin chains ${\rm S}=1/2$, noticing that
integrable spin-${\rm S}$ chains can be essentially treated along the same lines.
The higher Hamiltonian densities can be obtained by the iterative application of the boost operator $\hat{B}$,
\begin{equation}
\hat{H}^{(n+1)}\simeq [\hat{B},\hat{H}^{(n)}],\qquad
\hat{B}=\frac{1}{2\ii}\sum_{j}j\,\hat{\vec{S}}_{j}\cdot \hat{\vec{S}}_{j+1}.
\end{equation}
In close analogy to the isotropic Landau--Lifshitz magnet, the `second Hamiltonian flow' $H^{(3)}$
corresponds to the chiral three-spin interaction
\begin{equation}
\hat{h}^{(3)} = \hat{\vec{S}}_{j}\cdot (\hat{\vec{S}}_{j+1}\times \hat{\vec{S}}_{j+2}),
\end{equation}
whereas the Hamiltonian density of the `third flow' is supported on four adjacent lattice sites
\begin{equation}
\hat{h}^{(4)} = 2\hat{\vec{S}}_{j}\cdot \big(\hat{\vec{S}}_{j+1}\times (\hat{\vec{S}}_{j+2}\times \hat{\vec{S}}_{j+3})\big)
+2\hat{\vec{S}}_{j}\cdot \hat{\vec{S}}_{j+2}-4\hat{\vec{S}}_{j}\cdot \hat{\vec{S}}_{j+1}.
\end{equation}

Now we turn to the computation of the spin diffusion constant. The total contribution can be conveniently presented as a spectral
sum over quasi-particle excitations. The latter form an infinite tower of bound states made out of $s$ constituent magnons
carrying $s$ quanta of (bare) magnetization.
The spin diffusion constant associated with the $n$-th  Hamiltonian flow can be accordingly decomposed as
$\mathfrak{D}^{(n)}=\sum_{s\geq 1}\mathfrak{D}^{(n)}_{s}$, with contributions of individual quasi-particle species $s$
given by a closed-form expression \cite{NMKI2019}
\begin{equation}
\mathfrak{D}^{(n)}_{s} = \frac{1}{2}\int_{\mathbb{R}} \dd \theta\,n_{s}(\theta)(1-n_{s}(\theta))
\big|\varepsilon^{\prime\,(n)}_{s}(\theta)\big|
\left(\frac{\mu^{\rm dr}_{s}}{2\chi_h}\right)^{2},
\label{eqn:D_s}
\end{equation}
where the integration is taken over the range of the rapidity variable $\theta$,
$n_{s}(\theta)$ are Fermi occupation functions of the reference half-filled (i.e. $\expect{S^{z}}=0$) equilibrium background,
$\varepsilon^{\prime\,(n)}_{s}(\theta)$ denote the dressed values of the energy derivatives 
pertaining to the $n$-th Hamiltonian flow, $\mu^{\rm dr}_{s}$ are quasi-particles' dressed magnetic moments,
\begin{equation}
\mu^{\rm dr}_{s} \equiv \lim_{h\to 0} \frac{\partial^{2} m^{\rm dr}_{s}}{\partial h^{2}},
\end{equation}
and $\chi_h = \int \dd x \langle S^z(x) S^z(0)\rangle$ is the rescaled spin susceptibility at half filling.

The task at hand is to isolate the conditions under which $\mathfrak{D}^{(n)}$ becomes divergent.
It is sufficient to inspect the high-temperature limit of the grand-canonical Gibbs ensemble,
\begin{equation}
\hat{\rho} = \mathcal{Z}^{-1}\exp{[h\,\hat{S}^{z}]},\qquad \mathcal{Z}={\rm Tr}\,\hat{\rho},
\end{equation}
where closed-form expressions are available (see \cite{NMKI2019}) in the half-filled $ h\to 0$ limit. In particular, functions
\begin{equation}
n_{s} = \frac{1}{(s+1)^{2}}\sim \frac{1}{s^{2}},\qquad
\mu^{\rm dr}_{s} = \frac{1}{3}(s+1)^{2}\sim s^{2},
\end{equation}
become independent of the rapidity variable $\theta$. Furthermore, using the exact expressions
\begin{equation}
\varepsilon^{\prime\,(2)}_{s}(\theta)=8\theta(s+1)
\left[\frac{1}{(4\theta^{2}+s^{2})^{2}}-\frac{1}{(4\theta^{2}+(s+2)^{2})^{2}}\right],
\end{equation}
and
\begin{equation}
\varepsilon^{\prime\,(n)}_{s}(\theta)=
\frac{\partial^{n-2} \varepsilon^{\prime\,(2)}_{s}(\theta)}{\partial \theta^{n-2}} \qquad {\rm for}\quad n>2,
\end{equation}
we deduce that
\begin{equation}
\int \dd \theta\,| \varepsilon^{\prime\,(n)}_{s}(\theta)|\sim \frac{1}{s^{n}}.
\end{equation}
Based on this we conclude that the sum \eqref{eqn:D_s} is \emph{convergent} whenever $n\geq 4$.

We now take a closer look at anomalous cases $n=2,3$ with a divergent spin diffusion constant.
For this purpose we introduce the regularized diffusion constant,
\begin{equation}
\mathfrak{D}^{(n)}(s_{\star})=\sum_{s=1}^{s_{\star}}\mathfrak{D}^{(n)}_{s},
\end{equation}
where we have imposed a spectral cut-off $s_{\star}$ which integrates out the `heavy' quasi-particles. Now we can
essentially reiterate the dimensional analysis along the lines of ref.~\cite{GV2019}.
The key piece of information is the large-$s$ behavior of
the dressed velocities
\begin{equation}
v^{(n){\rm dr}}_{s}(\theta) = \frac{\partial \varepsilon^{(n)}_{s}(\theta)}{\partial p_{s}(\theta)},
\end{equation}
where $p_{s}(\theta)$ denote momenta of dressed excitations.
At large $s$, these satisfy the algebraic law (to be intended under integration over $\theta$)
\begin{equation}
v^{(n){\rm dr}}_{s}(\theta)\sim \frac{1}{s^{n-1}}.
\end{equation}
The associated `anomalous diffusive length'
\begin{equation}
x_{\star}=v^{(n){\rm dr}}_{s}t,
\end{equation}
can then be converted into the time-domain by comparing it to the \emph{time-dependent} diffusion constant $\mathfrak{D}^{(n)}(t)$ via
\begin{equation}
x^{2}_{\star} \sim \mathfrak{D}^{(n)}(t)t.
\end{equation}

As previously shown in \cite{GV2019}, for $n=2$ one makes the ansatz $\mathfrak{D}^{(2)}(t)\sim t^{\alpha}$
and deduces that $s_{\star}\sim t^{(1-\alpha)/2}$. Inserting this result back to $\mathfrak{D}^{(2)}(s_{\star})\sim s_{\star}$ and
comparing it to $\mathfrak{D}^{(2)}(t)\sim t^{\alpha}$ yields the superdiffusive exponent $\alpha=1/3$
(which translates into the dynamical exponent $z=2/(\alpha+1)=3/2$).

In the $n=3$ case, where
\begin{equation}
\mathfrak{D}^{(3)}(s_{\star})\sim \log{(s_{\star})},
\end{equation}
we instead plug in an ansatz $\mathfrak{D}^{(3)}(t)\sim [\log{(t)}]^{r(t)}$. The self-consistent value for $r$, which follows
from the dimensional analysis requires that $\lim_{t\to \infty}r(t)=1$.

We owe to point out the mismatch in comparison to ref.~\cite{Devillard1992} which, using a perturbative analysis at one-loop order,
predicts the logarithmic correction of the type $\mathfrak{D}^{(3)}_{\rm DS}(t)\sim [\log{(t)}]^{1/2}$.
In contrast, our conclusion follows from a non-perturbative calculation based on exact spectrum of thermally-dressed
quasi-particle excitations, but it also relies on a scaling analysis that could in principle fail to distinguish
different types of logarithmic terms.

\section{Landau--Lifshitz hierarchy}

We consider the Landau--Lifshitz hierarchy of commuting Hamiltonian flows
\begin{equation}
\vec{S}_{t_{n}} = F^{(n)}_{\rm LL} = -\vec{S}\times \frac{\delta H^{(n)}}{\delta \vec{S}},
\end{equation}
which includes the celebrated continuous isotropic Heisenberg ferromagnet
\begin{equation}
\vec{S}_{t_{2}} = F^{(2)} = \vec{S}\times \vec{S}_{xx}.
\end{equation}

Below we analyze the spin dynamics with aid of the Frenet--Serret apparatus, mapping the spin-field $\vec{S}\in S^{2}$
to a dynamical smooth curve in Euclidean space $\mathbb{R}^{3}$.
To each point on a curve, parametrized by its arclength $x$, we attach
a triad of orthonormal vectors $\{\vec{e}_{i}\}_{i=1}^{3}$, representing the tangent, normal and binormal vectors of the curve.
The local change of frame is then generated by a pair of $\mathfrak{so}(3)$ transformations,
\begin{equation}
(\vec{e}_{i})_{x} = \vec{\Omega}\times \vec{e}_{i},\qquad (\vec{e}_{i})_{t} = \vec{\omega}\times \vec{e}_{i},
\end{equation}
specified by the Darboux and angular-velocity vectors
\begin{equation}
\vec{\Omega} \equiv \tau \vec{e}_{1} + \kappa \vec{e}_{3},\qquad
\vec{\omega} = \sum_{i=1}^{3}\omega_{i}\vec{e}_{i},
\end{equation}
satisfying compatibility relation $\vec{\Omega}_{t}-\vec{\omega}_{x}=\vec{\Omega}\times \vec{\omega}$.
Identifying the spin-field with the tangent vector, $\vec{e}_{1}\equiv \vec{S}$, the time-evolution \eqref{eqn:abstract_EOM}
can be cast in the form $\vec{S}_{t} = (\vec{e}_{1})_{t} = \omega_{3}\vec{e}_{2} - \omega_{2}\vec{e}_{3}$, or
in terms of curvature and torsion as
\begin{equation}
\kappa_{t} = (\omega_{3})_{x}+ \tau \omega_{2},\qquad
\tau_{t} = (\omega_{1})_{x}-\kappa \omega_{2}.
\end{equation}
Therefore, we can express $\vec{S}_{xx} = -\kappa^{2}\,\vec{e}_{1}+\kappa_{x}\,\vec{e}_{2}+\kappa \tau\,\vec{e}_{3}$,
whence we deduce the angular velocities
\begin{equation}
\omega_{1} = \frac{\kappa_{xx}}{\kappa} - \tau^{2},\qquad
\omega_{2} = -\kappa_{x},\qquad
\omega_{3} = -\kappa \tau,
\end{equation}
and accordingly the Frenet--Serret equations
\begin{equation}
\kappa_{t_{2}} = -2\kappa_{x}\tau - \kappa \tau_{x},\qquad
\tau_{t_{2}} = \left(\frac{\kappa_{xx}}{\kappa}+\frac{\kappa^{2}}{2}-\tau^{2}\right)_{x}.
\end{equation}
These are also known as the Betchov-Da Rios equations \cite{DaRios,Barros1999} and govern the motion of a vortex filament in a viscous 
liquid.

The higher Hamiltonians $H^{(n\geq 3)}=\int \dd x\,h^{(n)}(x)$ can be constructed recursively \cite{Fuchssteiner1983},
\begin{equation}
h^{(n+1)}=\frac{1}{n}\,D^{-1}\left(\vec{S}_{x}\cdot D\,F^{(n)}_{\rm LL}\right),
\end{equation}
where $D\equiv \dd/\dd x$. In particular, the second (i.e. the third-order) flow is given
\begin{equation}
\vec{S}_{t_{3}} = F^{(3)}_{\rm LL} = -\vec{S}_{xxx} - \frac{3}{2}\big((\vec{S}_{x}\cdot \vec{S}_{x})\vec{S}\big)_{x},
\end{equation}
and is generated by the chiral interaction of the form
\begin{equation}
h^{(3)} = -\frac{1}{2}\vec{S}\cdot (\vec{S}_{x}\times \vec{S}_{xx}) = -\frac{1}{2}\kappa^{2}\tau.
\end{equation}
We shall in addition examine the third (i.e. fourth-order) flow
\begin{equation}
\vec{S}_{t_{4}} = F^{(4)}_{\rm LL} = \vec{S}\times \vec{S}_{xxxx}
+ \frac{5}{2}\big((\vec{S}_{x}\cdot \vec{S}_{x})\vec{S}\times \vec{S}_{x}\big)_{x},
\end{equation}
which corresponds to
\begin{equation}
h^{(4)} = \frac{1}{2}\left[\vec{S}_{xx}\cdot \vec{S}_{xx} - 5\big(h^{(2)}\big)^{2}\right]
= \frac{\kappa^{2}_{x}}{2}-\frac{\kappa^{2}}{2}\left[\frac{\kappa^{2}}{4} - \tau^{2}\right].
\end{equation}

The total integral of elastic energy density $\mathcal{E}\equiv h^{(2)}=\tfrac{1}{2}\kappa^{2}$ is a conserved quantity,
obeying the local conservation law
\begin{equation}
\mathcal{E}_{t_{n}} + \mathcal{J}^{(n)}_{x} = 0,
\label{eqn:fluxes}
\end{equation}
with flux densities
\begin{align}
\mathcal{J}^{(2)} &= h^{(3)} =  \kappa^{2}\tau,\\
\mathcal{J}^{(3)} &= \frac{1}{2}\kappa^{4}-\frac{1}{2}\kappa^{2}_{x} + \kappa \kappa_{xx},\\
\mathcal{J}^{(4)} &= -\frac{3}{2}\kappa^{4}\tau + 2\kappa^{2}\tau^{3} - \kappa^{2}\tau_{xx} -2\kappa\kappa_{x}\tau_{x}
-4\kappa\kappa_{xx}\tau +4\kappa^{2}_{x}\tau,
\end{align}
and so forth.

The angular velocities of the second flow $H^{(3)}$ read
\begin{align}
\omega_{1} &= \tau^{3} + \kappa^{2}\tau - \tau_{xx} - \frac{\kappa_{x}}{\kappa}\tau_{x} - 3\frac{\kappa_{xx}}{\kappa}\tau
-\frac{3}{2}\kappa \tau,\\
\omega_{2} &= 2\kappa_{x}\tau + \kappa \tau_{x},\\
\omega_{3} &= \kappa \tau^{2} - \frac{1}{2}\kappa^{3} - \kappa_{xx},
\end{align}
implying the following Frenet--Serret equations for the curvature and torsion
\begin{align}
\kappa_{t_{3}} + \left(\kappa_{xx}+\frac{1}{2}\kappa^{3}\right)_{x} + \frac{3}{2}\frac{(\kappa^{2}\tau^{2})_{x}}{\kappa} &= 0,\\
\tau_{t_{3}} + \left(\tau_{xx}+3\frac{(\kappa_{x} \tau)_{x}}{\kappa} + \frac{3}{2}\kappa^{2}\tau - \tau^{3}\right)_{x} &= 0.
\end{align}

\medskip

The above exact dynamical equations are still exact. The next step is to simplify them by taking into account
that on a large coarse-graining scale $\ell$ the curvature and torsion components of a hydrodynamically modulated soft mode
obey $\kappa \sim \tau \sim \mathcal{O}(1/\ell)$, which in effect allows to neglect the fluxes in Eq.~\eqref{eqn:fluxes}.
This means, in particular, that to the leading-order approximation
$\mathcal{E}\sim \mathcal{O}(1/\ell^{2})$ (and likewise $\kappa$) can be treated as constant and thus effectively decouple from
the torsion dynamics. Additionally, by dropping the dispersive terms in the equation for $\tau$,
we are left with the cubic Burgers' equation
\begin{equation}
\tau_{t_{3}} - (\tau^{3} + \ldots)_{x} = 0.
\end{equation}
 
The structure of the fourth flow $H^{(4)}$ is slightly more involved, and a lengthy calculation yields
\begin{align}
\omega_{1} &= \tau^{4} + \tau^{2}\left(-\frac{3}{2}\kappa^{2}-6\frac{\kappa_{xx}}{\kappa}\right)
+\tau\left(-12\frac{\kappa_{x}}{\kappa}\tau_{x}-4\tau_{xx}\right)+3\kappa^{2}_{x}+\frac{3}{2}\kappa \kappa_{xx}-3\tau^{2}_{x}
+\frac{\kappa_{xxxx}}{\kappa},\\
\omega_{2} &= -\kappa_{xxx} - \frac{3}{2}\kappa^{2}\kappa_{x}+3\kappa_{x}\tau^{2}+3\kappa \tau \tau_{x},\\
\omega_{3} &= \kappa \tau^{3} -\frac{3}{2}\kappa^{3}\tau - 3\kappa_{x}\tau_{x} - 3\kappa_{xx}\tau - \kappa \tau_{xx}
-\frac{5}{2}\kappa^{3}\tau.
\end{align}
In this case the Frenet--Serret equations are of the form
\begin{align}
\kappa_{t_{4}} &= -6\kappa^{2}\kappa_{x}\tau + 4\kappa_{x}\tau^{3} -4\kappa_{xxx}\tau - \frac{3}{2}\kappa^{3}\tau_{x}
+6\kappa\tau^{2}\tau_{x} - 6\kappa_{xx}\tau_{x} - 4\kappa_{x}\tau_{xx} - \kappa \tau_{xxx},\\
\tau_{t_{4}} &= (\tau^{4})_{x} + \frac{6}{\kappa^{2}}\left[\kappa_{x}\kappa_{xx}-\kappa^{3}\kappa_{x}-\kappa \kappa_{xxx}\right]\tau^{2}
+\frac{2}{\kappa^{2}}\left[6\kappa^{2}_{x}\tau_{x}-3\kappa^{4}\tau_{x}-12\kappa\kappa_{xx}\tau_{x}
-6\kappa\kappa_{x}\tau_{xx}-2\kappa^{2}\tau_{xxx}\right]\tau \nonumber \\
&+\frac{1}{2\kappa^{2}}\left[3\kappa^{5}\kappa_{x}+15\kappa^{2}\kappa_{x}\kappa_{xx}+5\kappa^{3}\kappa_{xxx}
-2\kappa_{x}\kappa_{xxxx}+2\kappa \kappa_{xxxxx}-24\kappa \kappa_{x}\tau^{2}_{x}-20\kappa^{2}\tau_{x}\tau_{xx}\right].
\end{align}
leading to, after repeating the above logic, a quartic Burgers' equation
\begin{equation}
 \tau_{t_{4}} - (\tau^{4} + \ldots)_{x} =0.
\end{equation}

\section{Additional numerical data}

\begin{figure}[htb]
\centering
\includegraphics[width=0.4\columnwidth]{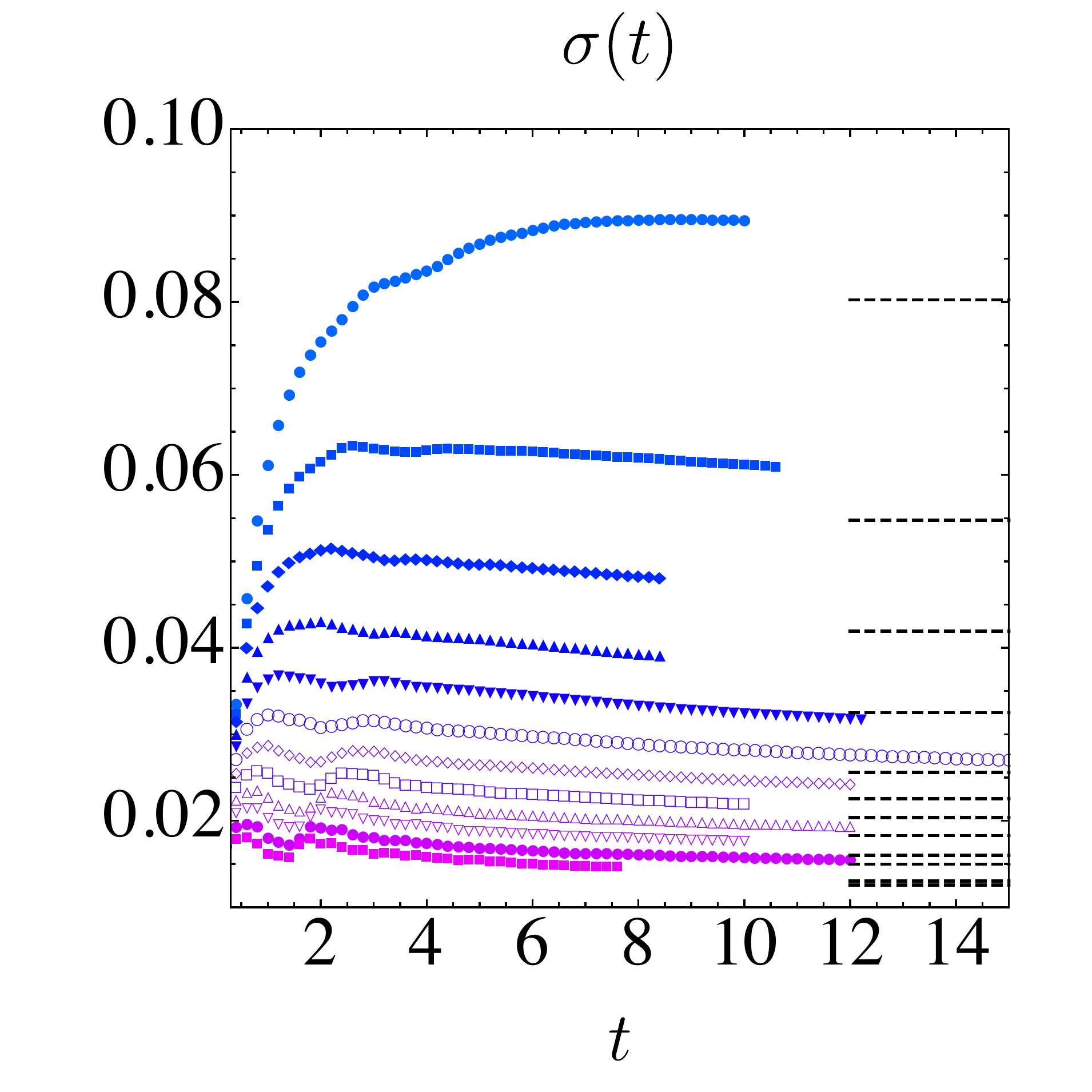}
\caption{Spin conductivity $\sigma(t)=\int_0^{t} dt'\sum_x \langle j_x(t') j_0 \rangle_T$ of the \textit{quantum} spin-1 chain $\hat{H}_\delta$ as function of time, shown at temperature $T=10$ and for values of $\delta$ ranging from $\delta=0.25$ (light blue) to $\delta=3$ (purple) with spacing of $0.25$, computed with a tDMRG algorithm. The dashed lines on the right of the plot show the fitted time-asymptotic limit $\lim_{t \to \infty} \sigma(t) = \chi \mathfrak{D}/T$, with $\chi$ static spin susceptibility.
The data is compatible with diffusion constant $\mathfrak{D}$ diverging as the isotropic point $\delta =0$ is approached.}
\label{fig:sigmabeta01}
\end{figure}

\begin{figure}[t]
\centering
\includegraphics[width=0.4\columnwidth]{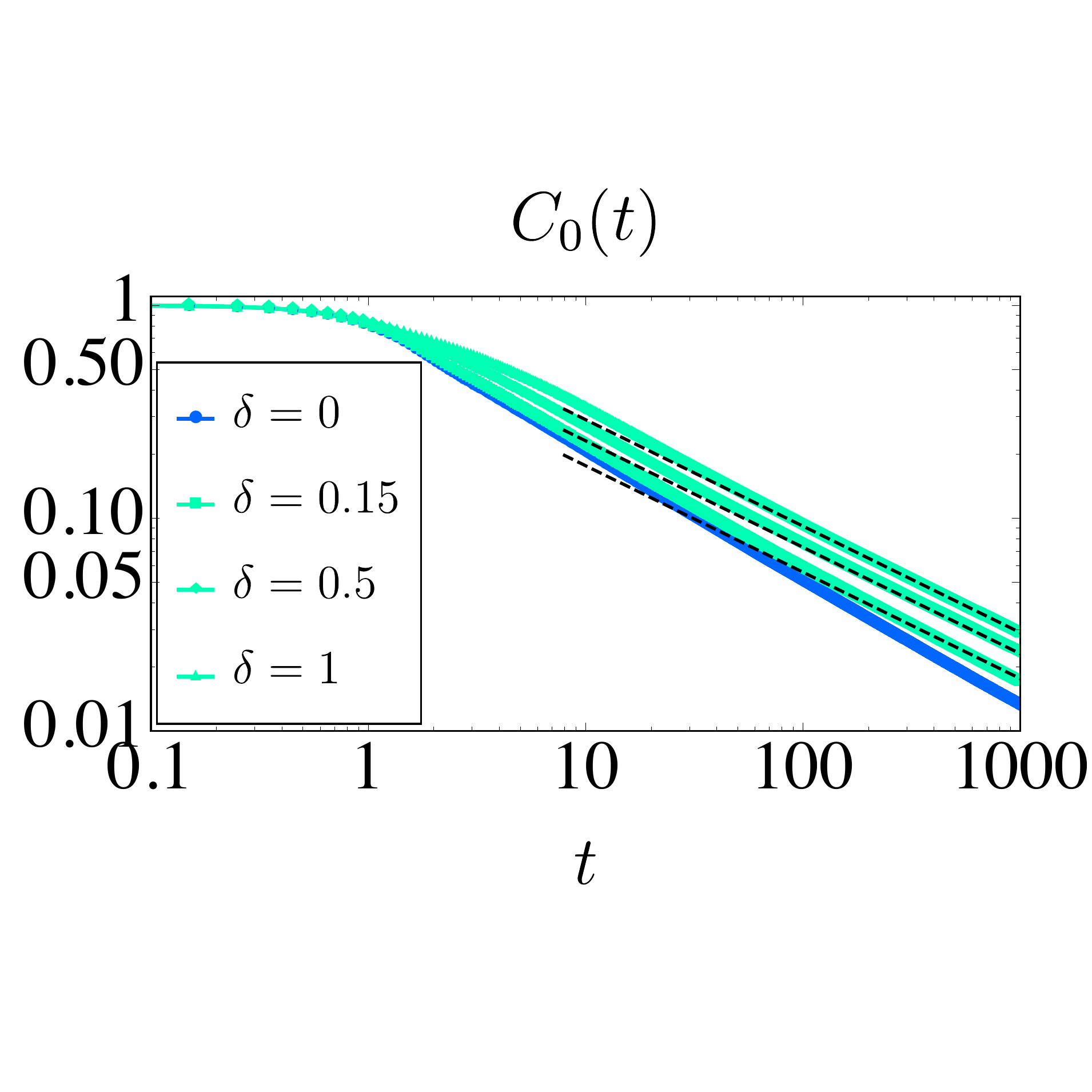}
\caption{Spin autocorrelation ${C}_0(t)= \langle S_{L/2}^z(t) S_{L/2}^z(0) \rangle$ in the classical ${H}_\delta$ chain at infinite temperature for different anisotropies $\delta$ shown on the log-log scale.
The decay of the autocorrelation crosses over from a fast decay to normal diffusive scaling $C_0(t)\simeq (2 \pi \mathfrak{D} t)^{-1/2}$ (dashed black lines) with spin diffusion constant $\mathfrak{D}$ diverging as $\delta \to 0^+$.  }
\label{fig:sigmabeta01}
\end{figure}

\begin{figure}[t]
\centering
\includegraphics[width=0.4\columnwidth]{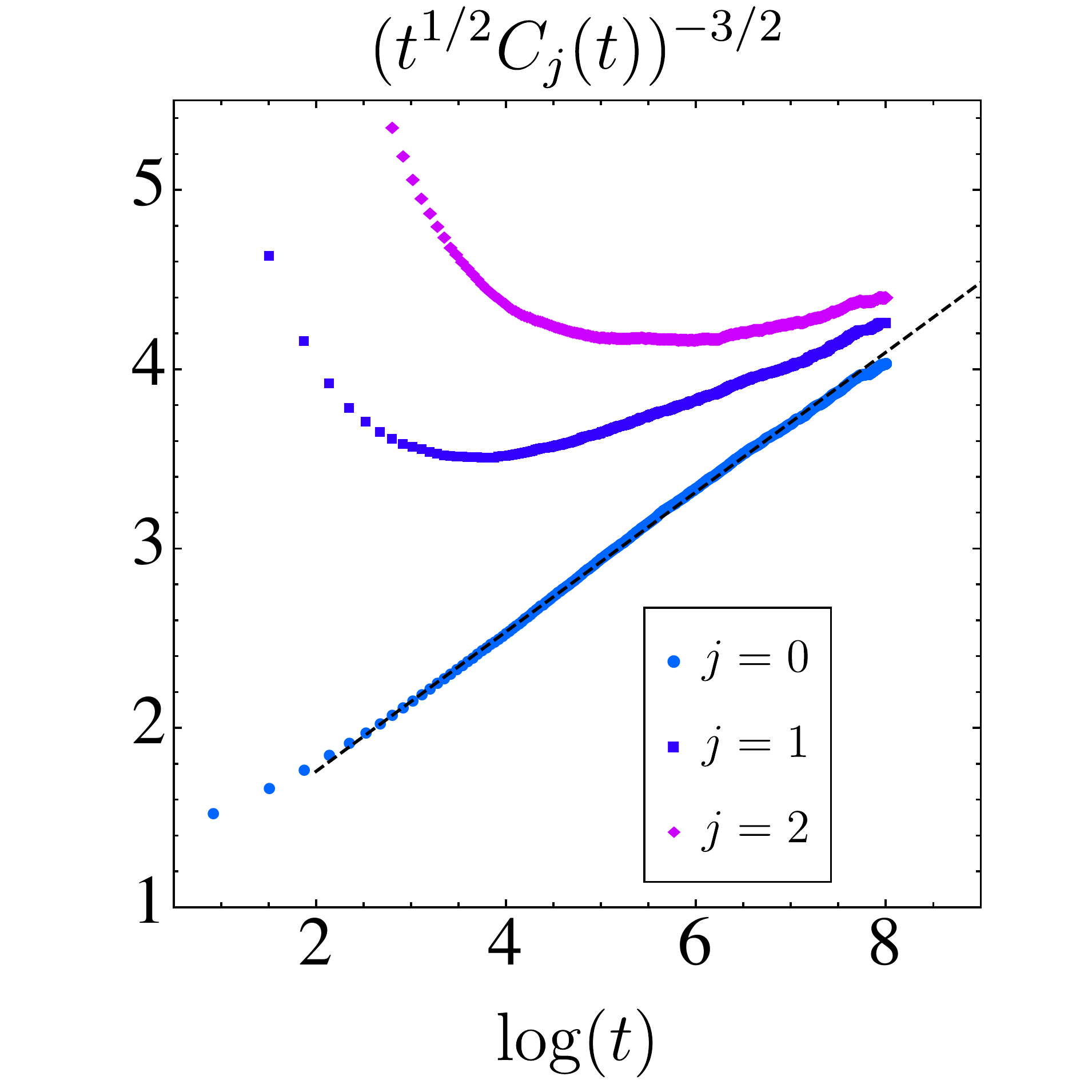}
\caption{Spin autocorrelation ${C}_j(t)= \langle S_{L/2+j}^z(t) S_{L/2}^z(0) \rangle$ in the anisotropic classical spin chain 
${H}_\delta$ at infinite temperature, shown for different value of $j$. The data is consistent with convergence towards anomalous 
diffusion law $\langle S_{L/2}^z(t) S_{L/2 + j}^z(0) \rangle \sim t^{-1/2} (\log t)^{-3/2}$.}
\label{fig:sigmabeta01}
\end{figure}

\begin{figure}[t]
\centering
\includegraphics[width=0.4\columnwidth]{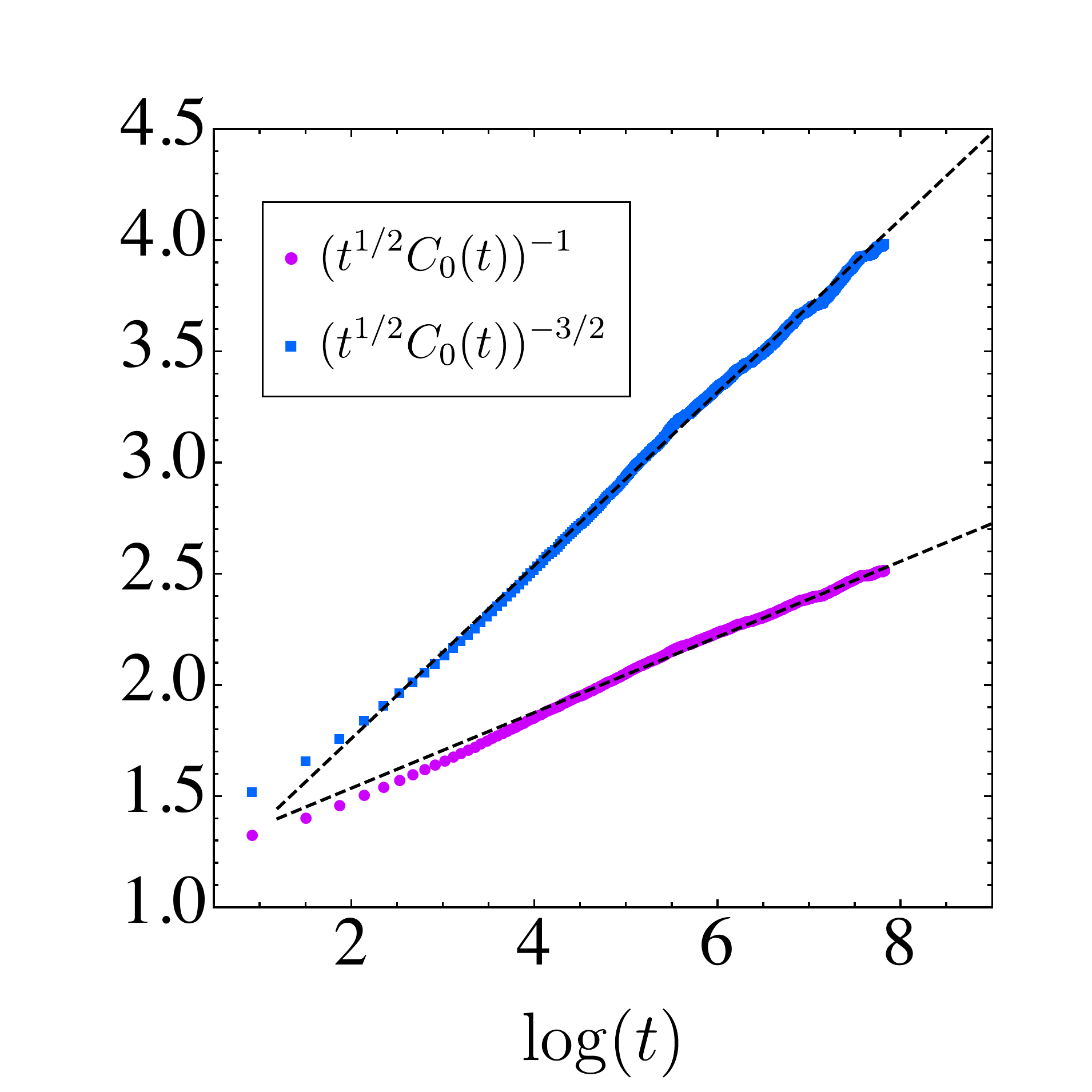}
\caption{Plot of the rescaled spin autocorrelation function $[t^{1/2}C_{0}(t)]^{-1}$ and $[t^{1/2}C_{0}(t)]^{-3/2}$  as
function of $\log{(t)}$, computed in the (non-integrable) classical Heisenberg chain. The dashed curves are fitting lines with $\log{(t)}$. The numerical data is unable to reliably distinguish between the decay $C_{0}(t) \sim t^{-1/2}(\log t)^{-1}$ and
$C_{0}(t) \sim t^{-1/2}(\log t)^{-2/3}$, despite the latter is a slightly better fit.}
\label{fig:logtdecay2}
\end{figure}

\end{document}